\documentclass{aa501}
\usepackage{natbib,graphics,epsf,psfig}
\bibpunct{(}{)}{;}{a}{}{,}
\begin{document}

%
   \title{Equilibrium abundances in hot DA white dwarfs \\
     as derived from self-consistent diffusion models}

   \subtitle{I. Analysis of spectroscopic EUVE data
     \thanks{Based on observations made with the EUVE Satellite.} 
            }

   \author{S.L. Schuh
          \inst{1}
          \and
          S. Dreizler\inst{1}
          \and
          B. Wolff\inst{2}
          }

   \offprints{S.L. Schuh}

   \institute{Institut f\"ur Astronomie und Astrophysik (IAAT), Universit\"at
              T\"ubingen, D--72076, T\"ubingen, Germany,
              \email{schuh@astro.uni-tuebingen.de, dreizler@astro.uni-tuebingen.de}
         \and
             Institut f\"ur Theoretische Physik und Astrophysik,
              Universit\"at Kiel,
              D--24098 Kiel, Germany, \\
              \email{wolff@astrophysik.uni-kiel.de} 
             }

   \date{Received date; accepted date}
   \titlerunning{Equilibrium abundances in hot DA white dwarfs I.}

   \abstract{
     We present the first analysis of an EUV selected sample of hot DA
     white dwarfs using a new type of atmospheric models. These models take
     into account the interplay between gravitational settling and
     radiative acceleration to predict the chemical stratification from an
     equilibrium between the two forces while self-consistently solving for
     the atmospheric structure. In contrast to atmospheric models with the
     assumption of chemical homogeneity, the number of free parameters in
     the new models is reduced to the effective temperature and surface
     gravity alone. The overall good reproduction of observed EUV spectra
     reveals that these models are able to describe the physical conditions
     in hot DA white dwarf atmospheres correctly. A comparison with
     previous analyses highlights the improvements as well as the limits of
     our new models.
     }
   
   \maketitle

%

\section{Introduction}
\label{sec:introduction}
White dwarf (WD) atmospheres exhibit a quasi--mono--elemental chemical
composition. They are either practically pure hydrogen or pure helium, with
the hydrogen rich sequence being called DA and the helium rich sequence
being summarized as non--DA.  In the following, we will restrict our
discussion to DA white dwarfs, even though similar physics applies to
non--DAs. The basic mechanism for purification of white dwarf envelopes
was first explained by \citet{schatzman:49}: The strong gravitational
field (typically, $\log{g}=8$ in cgs units) on the surface of WDs yields a
steep pressure gradient. Pressure driven diffusion separates the elements
according to their atomic weight, which is generally referred to as
gravitational settling or sedimentation.
\par
In \emph{hot} WDs (an approximate effective temperature range starts from
$T_{\rm eff}\approx 40,000\,{\rm K}$ upwards), however, this diffusion
process is disturbed by the radiation pressure that elements experience,
depending on the overlap of their wavelength--dependent opacity with the
maximum spectral flux. 
The overlap of the maximal spectral intensity with the region of maximal
opacity yields a considerable radiative acceleration on heavier elements in
WD atmospheres, provided the luminosity is high enough
\citep{vauclair:79,fontaine:79}. In the considered temperature range the
luminosities are indeed still high (i.e. L/L$_\odot > 1$).
This is sufficient to sustain photospheric elements other than hydrogen with
abundances no higher than $10^{-4}$ relative to hydrogen. In \emph{optical}
spectra of hot white dwarfs, these trace pollutants become directly visible
only when great effort is put in the detection of individual lines
\citep{dupuis:00}.  Through resonance lines in the UV and even more through
the sheer number of lines in the EUV, however, these spectral ranges are
strongly affected. The flux deficit caused by the additional opacity in the
EUV as compared to pure hydrogen atmospheres, hinted at already by earlier
data \citep{mewe:75,hearn:76,lampton:76,margon:76,shipman:76}, became more
and more evident as the quality of observations improved with the
HEAO(Einstein), EXOSAT and ROSAT satellites
\citep{kahn:84,petre:86,jordan:87,paerels:89,barstow:93b,jordan:94,wolff:96}.
While one EXOSAT \citep{vennes:89} and various ROSAT observations revealed
the opacity to be mainly due to absorbers other than helium, observations
with EUVE made possible a more detailed investigation of the nature of
absorbers.
\par
The most recent analyses of hot white dwarf EUVE spectra have been
presented by \citet{barstow:97a} and \citet{wolff:98a}.  Using the latest
non-LTE and LTE atmosphere models, both groups derived a metal mix scaling
factor for each object in their respective WD sample.  Barstow et~al. use
this single scaling factor per object to adjust metal abundances predicted
for its combination of effective temperature and surface gravity by
\citet{chayer:95a,chayer:95b}, whereas \citeauthor{wolff:98a} use relative
metal abundances which they have derived for the standard star G\,191--B2B
as a typical metal mix and give the appropriate scaling factor that they
call \emph{metallicity} for each of their sample objects.
\par
In the models used, the distribution of the metals was assumed to be
homogeneous, i.e. abundances were treated as being the same in all
atmospheric depths. But as the effects of sedimentation and radiative
levitation lead to a chemical gradient, fitting the emergent flux of real,
stratified WD atmospheres with homogeneous atmosphere models has its
limits, and true consistency cannot be achieved. Therefore, attempts were
made to include the impact of a chemical gradient via an ad hoc
stratification: For iron in G\,191--B2B \citep{barstow:99}, which could
successfully reproduce the EUVE spectrum of G\,191--B2B, and for nitrogen
in RE\,J1032+535 \citep{holberg:99b}.  
\par
As discussed above, the basic physical mechanisms leading to the
stratification have long been known, so their actual modelling appears
preferable to an ad hoc approach.  \citet{chayer:95a,chayer:95b} were the
first to calculate an extensive model grid for white dwarfs predicting the
chemical stratification assuming equilibrium between gravitational settling
and radiative levitation. An improvement has now been achieved through the
introduction of self-consistent non-NLTE model atmospheres which account
for the coupling between chemical stratification and the radiation field
\citep{dreizler:99b}. The successful application of these models to the
EUVE spectrum of G\,191--B2B \citep{dreizler:99a} has motivated a more
systematic analysis of a larger sample of hot DA white dwarfs which we will
present here.
\par
The first paper on this investigation exploits EUVE spectra, further papers
will deal with the UV and optical wavelength ranges and give detailed metal
abundance pattern predictions. The availability of EUVE spectra defines, as
in \citet{wolff:98a}, the overall sample, which is more thoroughly
introduced in Sect.~\ref{sec:observations}. The observations and particular
considerations concerning the interstellar medium are shortly described
there as well. Sect.~\ref{sec:models} outlines the principle of the model
calculations and focuses on the characteristics of the model grid computed
for comparison with the sample on hand. The methods and results from
fitting the new class of theoretical spectra to the observations are
presented in Sect.~\ref{sec:matching}. What implications these results
involve, and why further efforts in this direction are both justified by
the present success and at the same time necessary due to remaining
puzzles, is being discussed in Sect.~\ref{sec:discussion}.

%

\section{EUVE: Observations of hot DA white dwarfs}
\label{sec:observations}
Generally speaking, the maximum spectral flux of hot WDs lies in the
extreme ultraviolet (EUV). The fact that the main opacity of heavy elements
at these temperatures tends to lie in the same range means that radiative
acceleration can sustain traces of them in the atmosphere, but it also
implies that the EUV is the best suited spectral range for the detection of
these elements. Consequently, we start by surveying EUVE spectra, even
though its instrument's spectral resolutions are not high enough to
identify individual lines. Identification element by element is, however, not a
necessary requirement with the new models: They impose no need for the use
of a pre-defined metal mix, nor do they require the adjustment of 
e.g. a scaling factor, as the abundances are not free parameters but
are derived from an equilibrium condition (see Sect.~\ref{sec:models}). 
As the EUV flux is very sensitive to the chemical composition of the
photosphere, a comparison in this regime is extremely useful to
test the predictions of the diffusion models. Interpreting the repeated
analysis as a test case for the models, it is reasonable to
start out by following previous investigations as closely as possible. For
this reason, we chose to use the 26 DAs already analyzed by \citet{wolff:phd}. 
\subsection{The EUV selected DA sample}
\label{subsec:sample}
This sample comprises all hot ($T_{\rm eff}>40,000\,{\rm K}$) DA white
dwarfs with EUVE observations, which are mostly
taken from the EUVE public archive (now available at the multimission data
archive at STScI), and a few from own observations. The data sets and 
the reduction procedures applied are described elsewhere
\citep{wolff:98a,wolff:phd}; those publications also show the reduced
spectra in the form used here, i.e. flux calibrated SW, MW and LW data
combined to yield a single spectrum. 
\par
The effective temperatures of the sample stars lie above $T_{\rm
  eff}=40,000\,{\rm K}$, where photospheric metal abundances still are large enough
to be detectable at all. Above $T_{\rm eff}=70,000\,{\rm K}$, mass loss effects
might start to disturb the expected equilibrium between gravitational and
radiative acceleration \citep{unglaub:98a}, but the sample contains no
stars hotter than that. \citet{wolff:98a} have grouped the
objects into four different categories. The first set consists of
G\,191--B2B like objects, where the high metallicity yields a steep flux
drop towards short wavelengths, beyond the interstellar \ion{He}{ii}
absorption edge. The second group comprehends GD\,246 and similar objects
which contain less photospheric metals, but still more than those in the
third group whose spectra Wolff et~al. could fit with pure hydrogen
atmospheres. A fourth rather inhomogeneous set compiles the remaining
objects. For details, especially on the effective temperature, surface
gravity and metallicity region in parameter space covered by the
different groups, we refer once more to \citet{wolff:phd}.
\subsection{Treatment of the ISM}
\label{subsec:ism}
As briefly remarked on before, the interstellar medium can considerably
attenuate the stellar flux. For the EUV, the most important features are
the \ion{H}{i}, \ion{He}{i} and \ion{He}{ii} bound-free ground state
absorptions with edges at 911.7\,\AA, 504.3\,\AA\ and 227.8\,\AA,
respectively. Again, this is being accounted for in the same way as in
\citet{wolff:98a}: The effect of given interstellar column densities on the
theoretical fluxes is modelled and applied following \citet{rumph:94}.  The
derivation of \ion{H}{i} column density depends on the effective
temperature of the model used, and the \ion{He}{i} and \ion{He}{ii} column
densities on its absorber content. Due to the interweavement of $T_{\rm
eff}$ and N(\ion{H}{i}) and the fact that the Lyman edge lies outside the
range observed by EUVE, \citet{wolff:98a} have started from optically
determined effective temperatures \citep[by][]{Finley:1997}, and
additionally have retained surface gravities from that analysis unchanged.
In the present approach, we start from the effective temperatures from
\citet{wolff:phd} and adopt his \ion{H}{i} column densities unalteredly
whenever possible.
\par
Even though the \ion{He}{i} and \ion{He}{ii} column densities have to be
constantly modified during the fitting procedure depending on the assumed
photospheric parameters, all interstellar contributions will in the
following be treated as a secondary effect on the stellar light that can be
dealt with independently of the analysis of the source itself, disregarding
the fact that an appropriate correction can only be made \emph{after} a set
of photospheric parameters has been adopted.  As the interstellar
absorption is more important at longer wavelengths while the absorption by
photospheric metals increases towards shorter wavelengths, the different
contributions can partly be separated.

%

\section{Equilibrium ansatz: Chemically stratified model atmospheres}
\label{sec:models}
Under the assumptions that no processes competing with diffusion and
levitation are present \emph{and} that diffusion timescales are short,
the distribution of trace elements in a
stellar atmosphere can be expected to take on an equilibrium state, as
proposed by \citet{chayer:95a,chayer:95b}. In such a situation, mean
diffusion velocities would be zero at any location in the atmosphere, due
to an exact balance of gravitational, radiative and electrical forces (when
neglecting thermal and concentration diffusion). A more detailed
formulation of this condition for polluting elements~$(i)$ in a hydrogen (or
helium) plasma~$(1)$ can be found elsewhere \citep{dreizler:99b}. It yields,
in a short outline,
\begin{equation}\label{forces}
  \begin{array}{lll}
    m_1 F_1- m_i F_i & = & (A_i-A_1)m_{\mathrm p}g - (Z_i-Z_1)eE \\
    &   & - A_im_{\mathrm p}g_{{\mathrm{rad}},i}\\
    & = & 0,\\ 
  \end{array}
\end{equation}
where $m_i F_i$ are the forces acting on element $(i)$, $A_i$ are the
atomic weights, $Z_i$ are the mean electrical charges, 
E is the electrical field caused by the separation of electrons and ions,
$m_p$ is the proton mass, and $g_{{\mathrm{rad}},i}$ is the radiative
acceleration acting on element~$(i)$.
Using hydrostatic equilibrium and charge conservation, this leads to the
simple term
\begin{equation}\label{geffeqgrad}
  g_{\mathrm{rad},i}=g_{\mathrm{eff},i}\, .
\end{equation}
One immediately recognizes that requiring diffusion velocities to be zero is
indeed the same as asking for the effects of gravitational settling and
radiative levitation to cancel each other. Knowing that the above
quantities stand for
\begin{equation}\label{geffisgradis}
  \begin{array}{lll}
    g_{\mathrm{eff},i} :&=&\left(1-\frac{A_1(Z_i+1)}{A_i(Z_1+1)}\right)g\\ \\
    g_{{\mathrm{rad}},i}&=&\frac{1}{\rho_i}\frac{4\pi}{c}\int_0^\infty
    \kappa_{\nu,i}H_{\nu}{\mathrm{d}}\nu
  \end{array}
\end{equation}
where $\rho_i$ is the mass density of the element $(i)$, $\kappa_{\nu,i}$ is
the frequency dependent absorption coefficient which includes all
contributions of this element at the frequency~$\nu$, and $H_{\nu}$ is the
Eddington flux, it furthermore becomes clear that
$g_{\mathrm{rad},i}=g_{\mathrm{eff},i}$ can only be fulfilled by fixed
values for ${\rho_i}$. This equilibrium solution for ${\rho_i}$ is depth
dependent. For a plane parallel model atmosphere in hydrostatic
equilibrium this means that equilibrium abundances are solely determined by
the photospheric parameters $T_{\rm eff}$ and $\log{g}$, in other words,
the photospheric abundances are no longer free parameters. Calculating the
equilibrium abundances inevitably leads to a chemically stratified
atmosphere.
\par
\citet{chayer:95a,chayer:95b} have presented extensive results of such
calculations for white dwarfs. Unfortunately, their predicted equilibrium
abundances were not quantitatively consistent with observations. This might
have been partly due to the fact that the observations were analyzed with
chemically homogeneous model atmospheres, meaning that homogeneous
abundances were compared to predictions for stratified atmospheres. It is
more probable yet an effect that arises because the feedback of the
modified abundances on the radiation field was not being accounted for.
\par
A self-consistent solution of the equilibrium condition and the atmospheric
structure has recently been presented \citep{dreizler:99b}. In addition to
this newly introduced coupling, it is also a novelty that the calculations
are being performed under non-LTE conditions. This has turned out to be a
crucial point \citep{schuh-mt:00,dreizler:01}. The code itself relies on an
iteration scheme that alternates between the determination of new
equilibrium abundances and the corresponding solution for the atmospheric
structure, which is described in \citet{werner:99a}.

\subsection{Scope of application}
\label{subsec:scope}
It remains to be justified why the white dwarfs in our sample will
presumably obey the presented equilibrium condition. Processes eligible to
disturb or prevent its adjustment can be mass loss, accretion from the
interstellar medium, convection or mixing through rotation. 
As explained by \citet{unglaub:98a}, mass loss would have the effect of
homogenizing a chemical stratification, but as has also been shown in the
same paper, mass loss rates drop below the critical limit
($10^{-16}$\,M$_\odot$/yr) for DAs cooler than $T_{\rm eff}=70,000\,{\rm
  K}$ so that this phenomenon should not occur in any of the sample stars. 
Calculations of \citet{macdonald:92} that treat the interaction between
accretion and the white dwarf wind reveal that WDs in our sample are not
affected by accretion. It is prevented since L$_{\rm WD}/L_\odot > 1$, so
that at least a critical mass loss rate of $3\cdot 10^{-18}$ to
$10^{-21}$\,M$_\odot$/yr is sustained.
\par
Convection would of course homogenize abundances,
too. However, convective instability in the outer layers occurs only below
$T_{\rm eff}=12,000\,{\rm K}$ for DAs, i.e. well below $T_{\rm
eff}=40,000\,{\rm K}$. Rotation could lead to a mixing through meridional
currents; but as most white dwarfs are very slow rotators
\citep{heber:97,koester:98}, this is not likely to interfere either.
\par
Even if it is left undisturbed, it takes an atmosphere a certain time to
reach an equilibrium state. Due to the high surface gravities, diffusion
time scales are of the order of months in the outer layers of white dwarfs
\citep[][based on diffusion coefficients by
\citealt{paquette:86b}]{koester:89iau}.  This is nothing but an instant
compared to evolution time scales ($\sim 10^{7}$\,yr), and this is the
reason why DA atmospheres in the effective temperature range between
$40,000\,{\rm K}$ and $70,000\,{\rm K}$ can be regarded to have their
photospheric abundances set to equilibrium values at any time of their
evolutionary stage.  In the temperature range above $40,000\,{\rm K}$ the
flux maximum still lies in the EUV, i.e. in the same range where multiple
lines of heavier elements provide substantial opacity, making radiative
levitation an efficient mechanism for sustaining absorbers. It should in
this context also be noted that the evaluation of the equilibrium
abundances does not include the contributions of thermal and concentration
diffusion, which have much longer time scales.

\subsection{The model grid}
\label{subsec:grid}

The model grid spans the $T_{\rm eff}$ and $\log{g}$ plane as suggested by
the results from \citet{wolff:phd}: The effective temperatures cover a
range from $38,000 - 70,000\,{\rm K}$ in $2,000 - 3,000\,{\rm K}$ steps,
the logarithmic surface gravity a range from $7.2 - 8.3$ in
$0.1 - 0.2~{\rm dex}$ steps. The grid is however not complete and
concentrates around the previously found parameter combinations for the
objects as listed in Table~\ref{tab:results}. 
We emphasize once more that the total
absorber content, i.e. the entire depth-dependent distribution of each
individual trace element, is solely defined by a models' [$T_{\rm
eff},\log{g}$] -- combination, which makes $T_{\rm eff}$ and $\log{g}$ the
\emph{only} free parameters. The elements included in the model
calculations are the same as those used by \citet{wolff:98a} for the
definition of their metal mix, which in turn are those that they were able
to identify in HST GHRS data of the standard G\,191--B2B. Other elements
were assumed to be present in abundances too low to significantly
contribute to the EUV opacity (however, stratification may well invalidate
this assumption). 
While other elements than those used in Wolff's and this analysis have been
identified in UV spectra of some of the sample stars, including G\,191--B2B
itself, the low abundances of $2.5 \cdot 10^{-8}$ for phosphorus and $3.2 \cdot
10^{-7}$ for sulfur, for example, as reported from the analysis of ORFEUS
spectra by \citet{vennes:96}, confirm that these element's opacities will
presumably not contribute significantly to the total EUV opacity. To
investigate the possible effects of several additional elements anyhow,
(which currently results in a dramatic increase of CPU time demand), further
analyses will have to be conducted (see also Sect.~\ref{subsec:perspective}). 
For this analysis, we aimed at being able to compare our results to
a previous analysis in the EUV, as mentioned before. 
Consequently, the list of
elements considered besides \element{H} reads
\element{He},
\element{C},
\element{N},
\element{O},
\element{Si},
\element{Fe} and
\element{Ni}.
Differences remain with regard to the number of atomic levels and with
regard to LTE/non--LTE population of these levels. Details for the model
calculations presented here are summarized in Table\,\ref{tab:levels}. Note
that so far, the model atoms used for the solution of the atmospheric
structure and for the solution of the equilibrium condition are identical,
which implies that the relatively high accuracy required for the correct
evaluation of the radiative acceleration slows down the atmospheric
structure calculation.
\par
The time it takes a model with the specified atomic data to converge
depends primarily on the stratification of the input model. For homogeneous
input models with abundances derived from the published metallicities as a
first guess, roughly 5x$10^{6}$ CRAY-CPU seconds are required, while using
converged stratified models with photospheric parameters that are somewhat
off the desired ones yields convergence after of the order of 5x$10^{5}$
CPU seconds.  Due to this particularly large CPU time demand per model, the
convergence criterion for the models was set to be a less than 5\% change
in the model flux, which means that the maximal relative correction at
\emph{any} frequency point is no larger than that limit, compared to the
result from the previous \emph{diffusion} iteration step.  Only single
lines (usually two or three \element{He} lines) are actually affected by
those larger variations, the continuum flux itself is much more exactly
determined and easily complies to our usual ``$<10^{-4}$'' criterion.
\par
The next generation of models will benefit from major improvements of the
model atmosphere code and the iteration scheme for the equilibrium
condition. Both will reduce the CPU time drastically.
\par
As to the uniqueness of a particular solution, tests indicate that the
results agree to within our error limits regardless of the extremely
different stratification of diverse start models, making a strong case at
least for the stability of the result. 
We therefore trust that, no matter what the initial conditions in the start
model are, the final abundances in the converged model will always be the
same.
\par
The emergent fluxes used for comparison with the observations have been
calculated from the model atmospheres on a frequency grid optimized for the
EUV spectral range. This grid covers $\lambda=50-700$\,\AA\ in detail and
contains more than 30,000 frequency points in total.

\begin{table}
  \caption[]{Summary of the model atoms. Super-levels (marked by
    $^*$) comprise hundreds to thousands of atomic levels. Line transitions
    between these super-levels (marked by $^+$) are composed of all
    transitions between the individual levels comprising the super-levels
    (see \protect\citealt{werner:99a} for details).}
  \label{tab:levels}
  \begin{tabular}{rcccccccc}
    \hline
    \noalign{\medskip}
    &\element{H~}
    &\element{He}
    &\element{C~}
    &\element{N~}
    &\element{O~}
    &\element{Si}
    &\element{Fe}
    &\element{Ni}\\
    \hline
   non--LTE &   &   &   &   &   &   &       &       \\
   levels   & 17& 54& 70& 50& 24& 25& 28$^*$& 29$^*$\\
   lines    & 56&106&218&114& 35& 53& 77$^+$& 44$^+$\\
    \noalign{\smallskip}
    \hline\hline
  \end{tabular}
\end{table}

\subsection{Characteristics of the models}
\label{subsec:characteristics}
The models show the general properties as expected from diffusion theory,
i.e. the overall trace element abundances decrease with lower $T_{\rm eff}$
and higher $\log{g}$ values. Accordingly, the emergent spectra approach the
flux distributions of those of pure hydrogen model atmospheres at
sufficiently evolved parameters on the cooling sequence.  The vertical
element distribution however does not generally follow the simple picture
of monotonously increasing local abundances with depth, but is strongly
affected by the respective radiative acceleration (see
\citealt{dreizler:99b}, \citealt{schuh-mt:00}, or \citealt{dreizler:01} for
exemplary graphical representations).  In particular, gradients may change
signs repeatedly, meaning this can be brought about by diffusion processes
alone, in contrast to other notions that additionally evoke mass loss
effects to explain local abundances that increase towards outer atmospheric
regions (e.g. \citealt{holberg:99b}). In a nutshell, even equilibrium
calculations yield rather non-trivial abundance distributions.
\par
As in previous calculations \citep{dreizler:99b,dreizler:99a}, the iron
group element abundance can reach solar values or more, with nickel being
at virtually the same level as the iron abundances.  Though their dominant
effect on the EUV opacity is partly due to the many lines these elements
exhibit, it is equally important that their abundances are, over a wide
parameter range, simply larger than those of other metals.  Of course these
two effects are tightly correlated, since more lines directly translate
into a stronger radiative acceleration, which in this case easily
compensates for the higher atomic weights and thus a stronger response to
gravity.
\par
This represents just a short overview of some general effects that can be
seen in the models; detailed presentation of depth dependent
stratifications for all elements included in the calculations will be
published separately.

%

%
\begin{figure*}[ht]
  \caption{EUVE spectrum of MCT\,2331$-$4731 with the best-fit model at 
    $T_{\rm eff} = 56,000\, {\rm K} $ and $\log{g} = 7.7$ overlaid
    (dashed). The photon flux is displayed on a logarithmic scale. 
}
  \label{fig:mct2331}
  \hbox{\hspace{0cm}  
    \epsfxsize=0.75\textwidth\epsffile{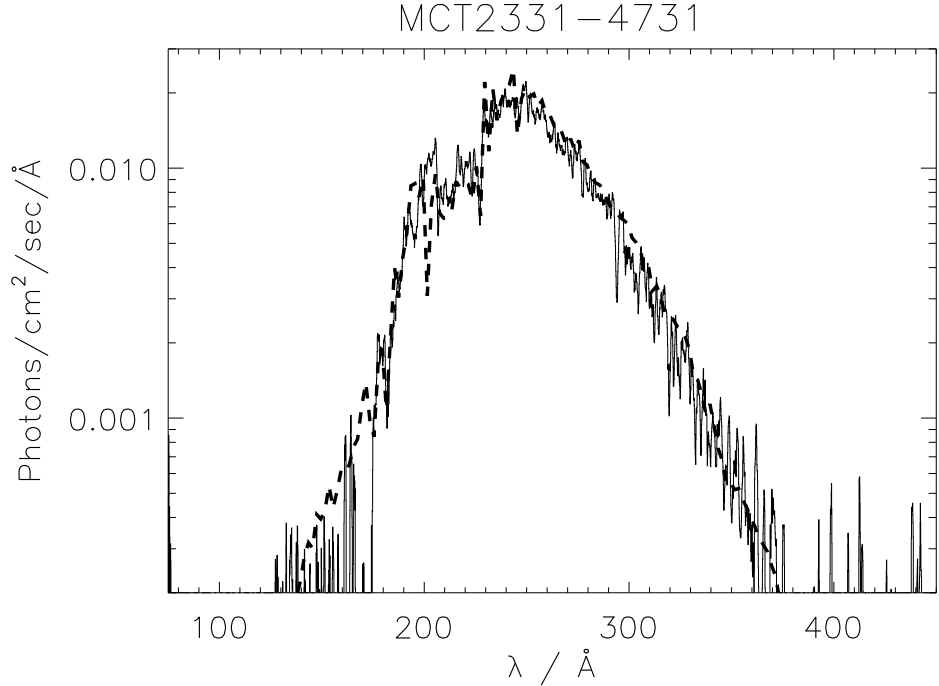} 
    }
\end{figure*}

%

%
\begin{figure*}[ht]
  \caption{EUVE spectra of the program stars with theoretical
    spectra overlaid (dashed), ordered by decreasing metal index $mi$. We
    only show 
    the spectral range with significant signal. Parameters of the models
    as well as interstellar column densities for hydrogen and helium can be
    found in Table\,\ref{tab:results}.
}
  \label{fig:fits}
  \hbox{\hspace{0cm}  
    \epsfxsize=0.26\textwidth \epsfbox{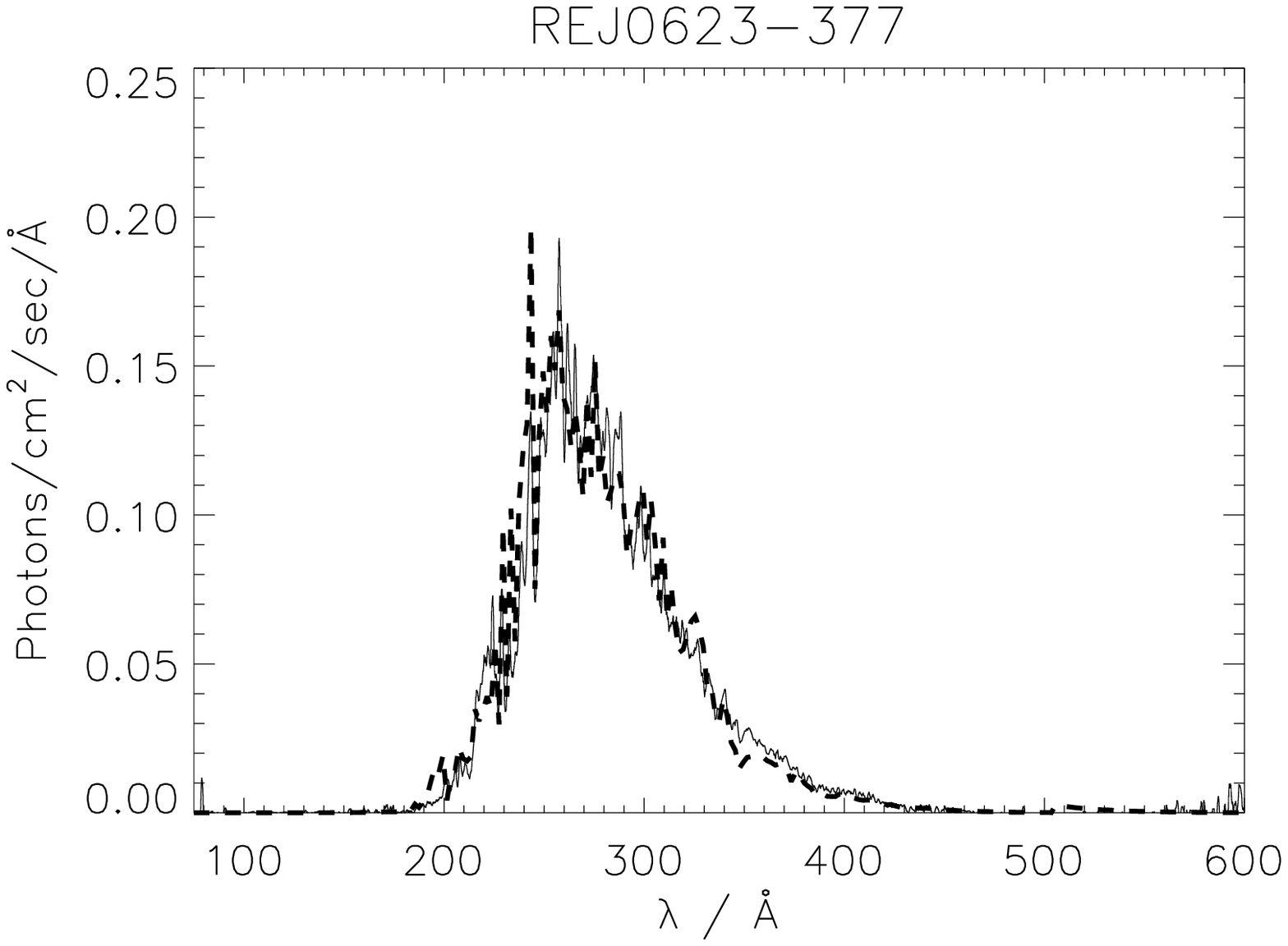}
    \hspace{-3mm}\epsfxsize=0.26\textwidth \epsfbox{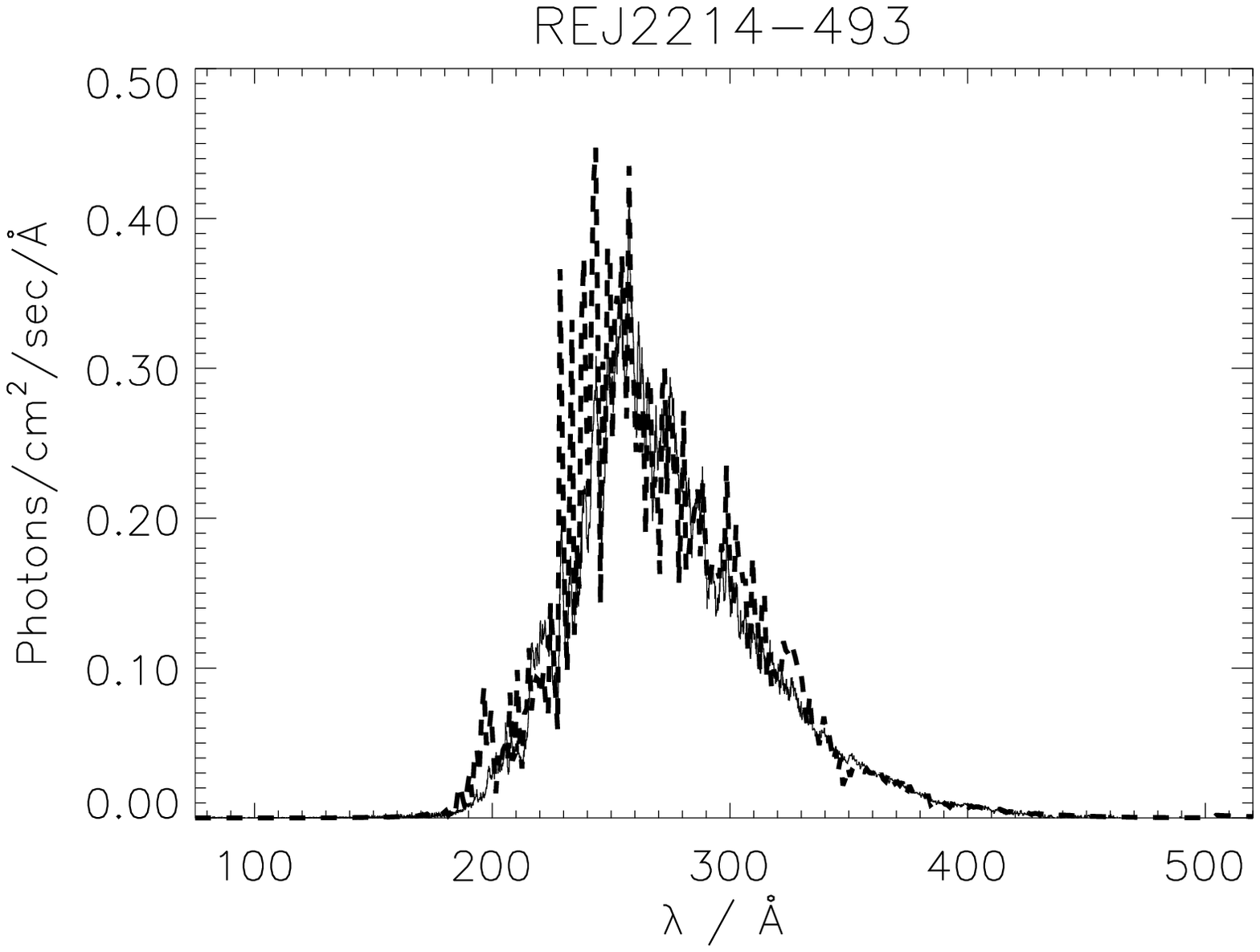}
    \hspace{-3mm}\epsfxsize=0.26\textwidth \epsffile{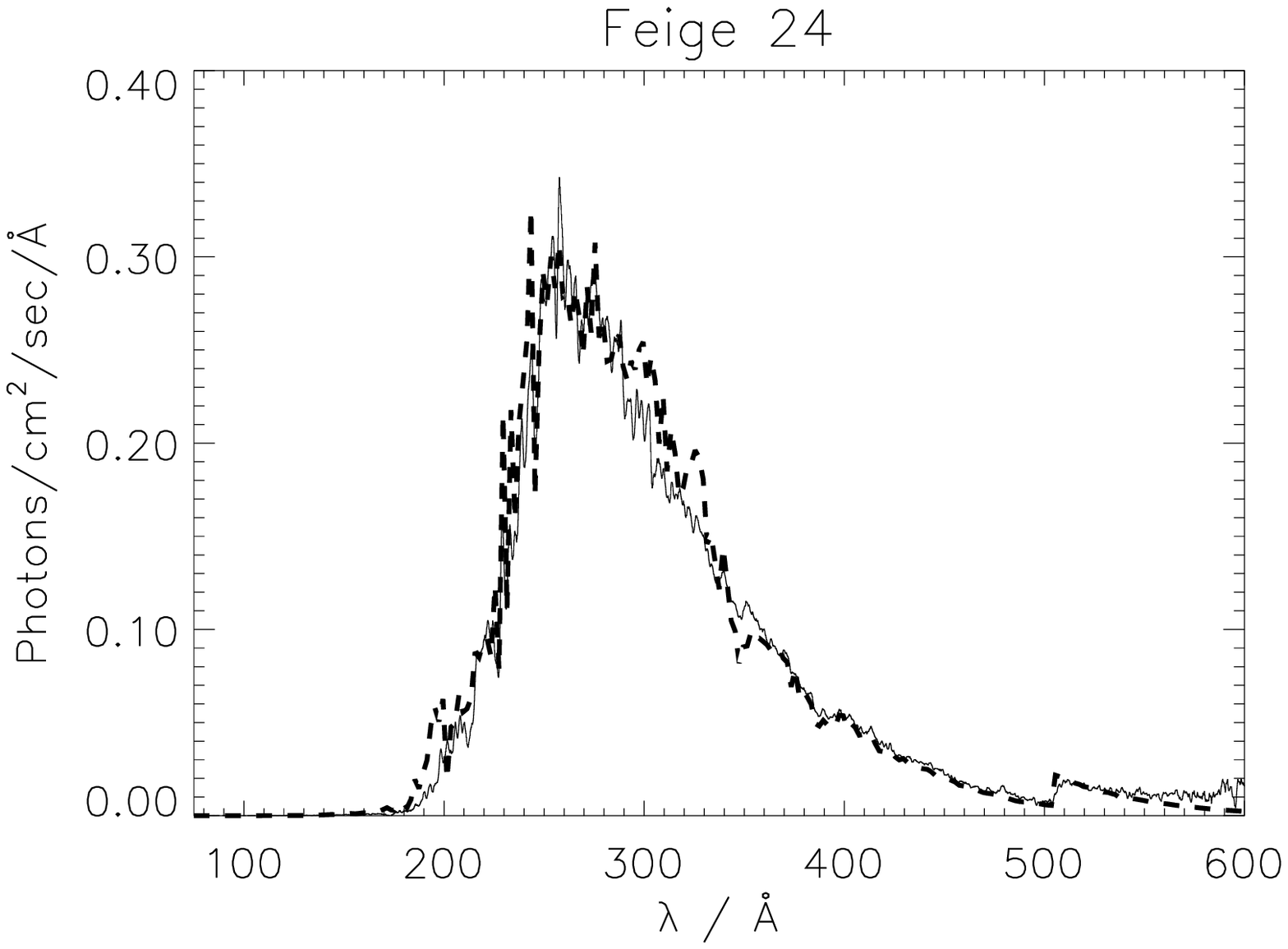}
    \hspace{-3mm}\epsfxsize=0.26\textwidth \epsffile{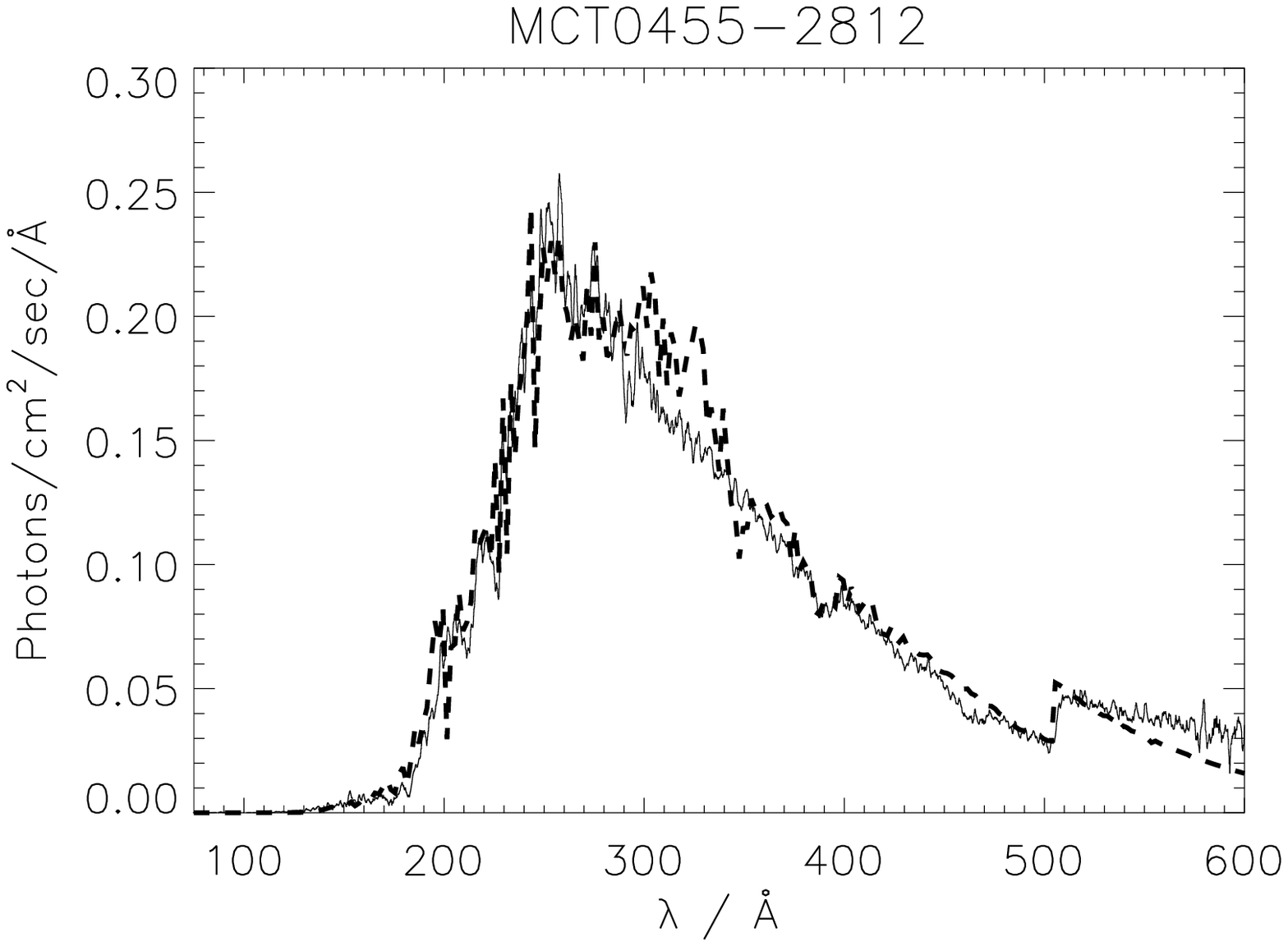}
    }
  \hbox{\hspace{0cm}  
    \epsfxsize=0.26\textwidth \epsffile{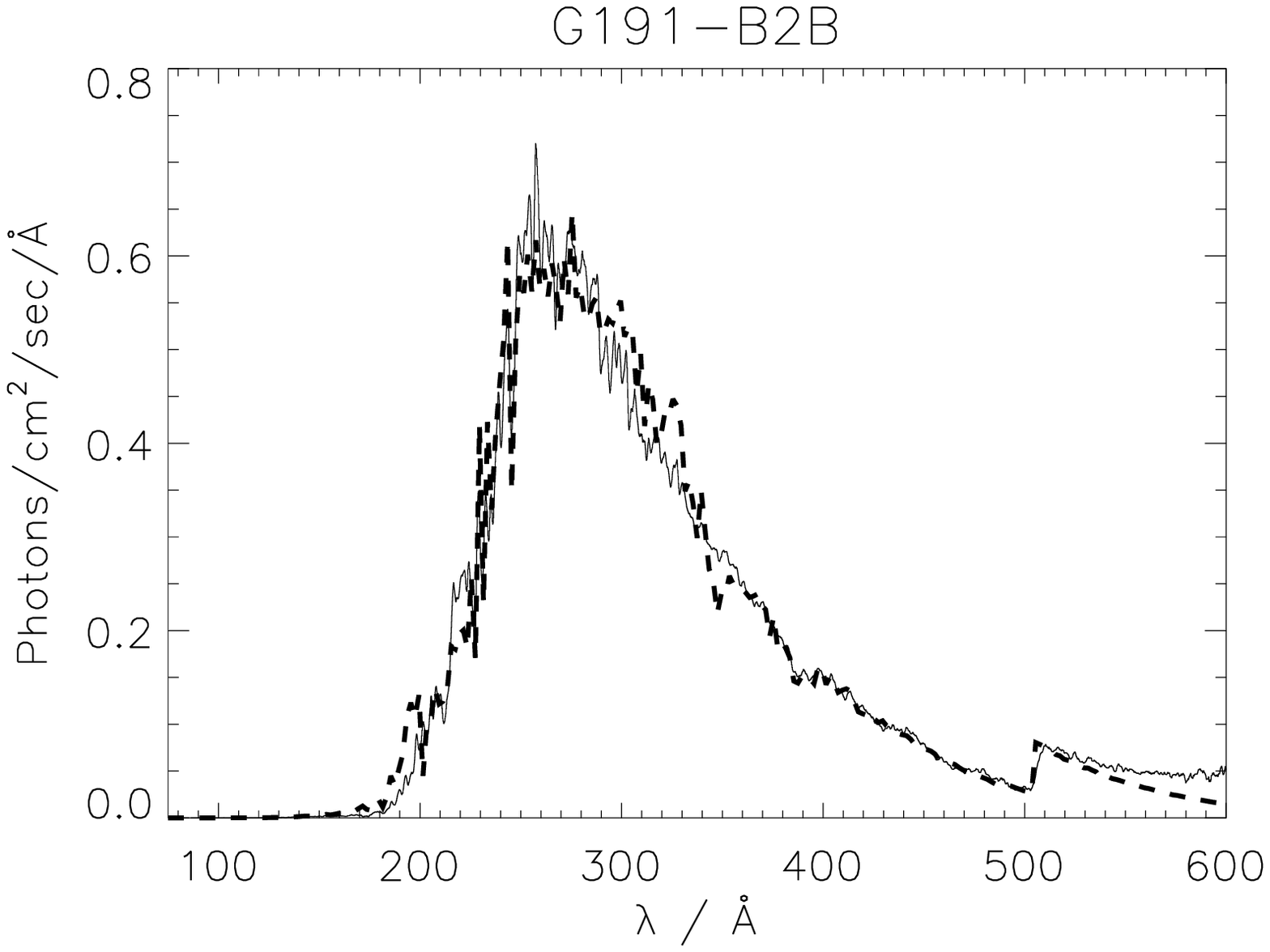}
    \hspace{-3mm}\epsfxsize=0.26\textwidth \epsffile{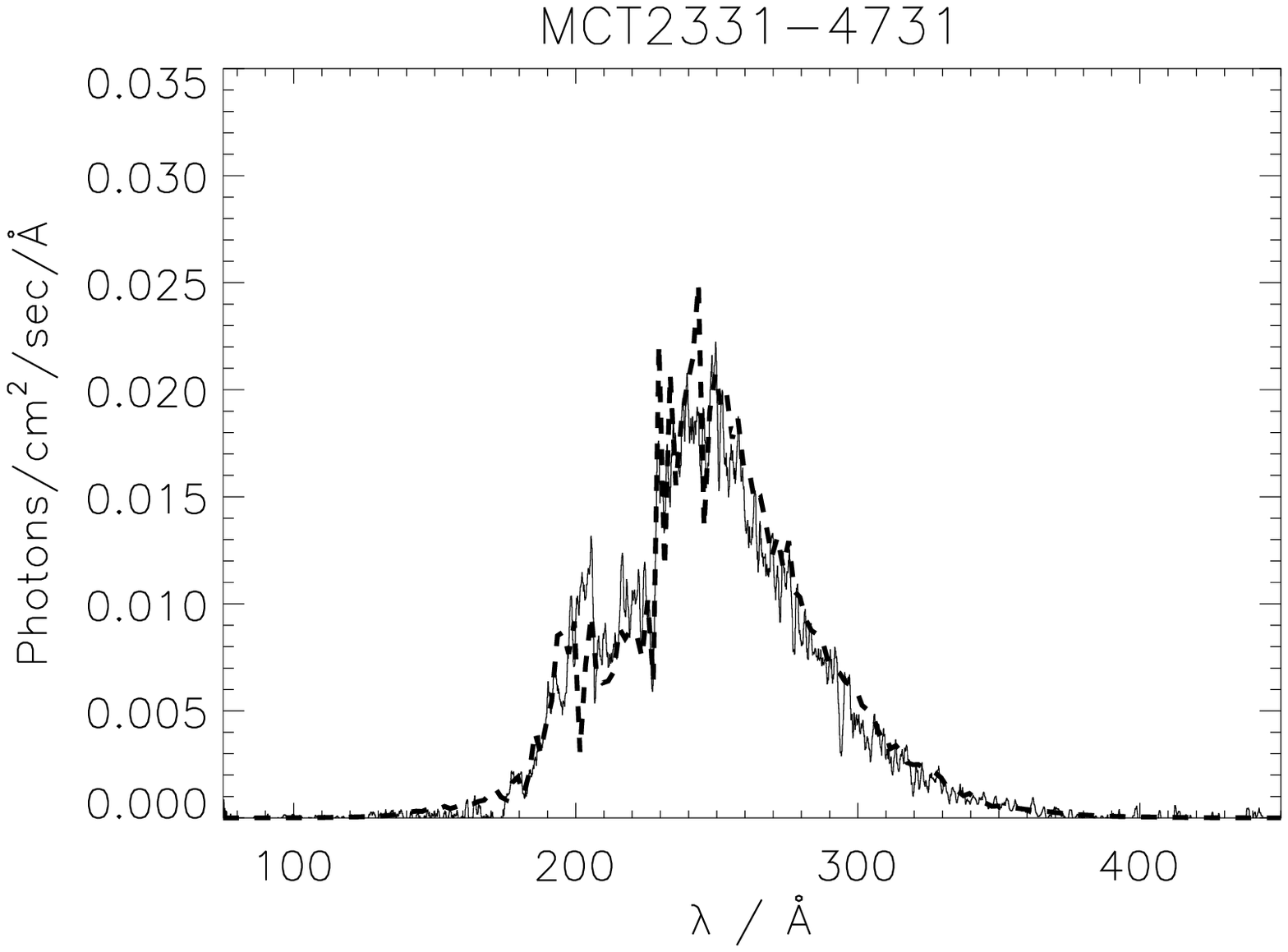}
    \hspace{-3mm}\epsfxsize=0.26\textwidth \epsffile{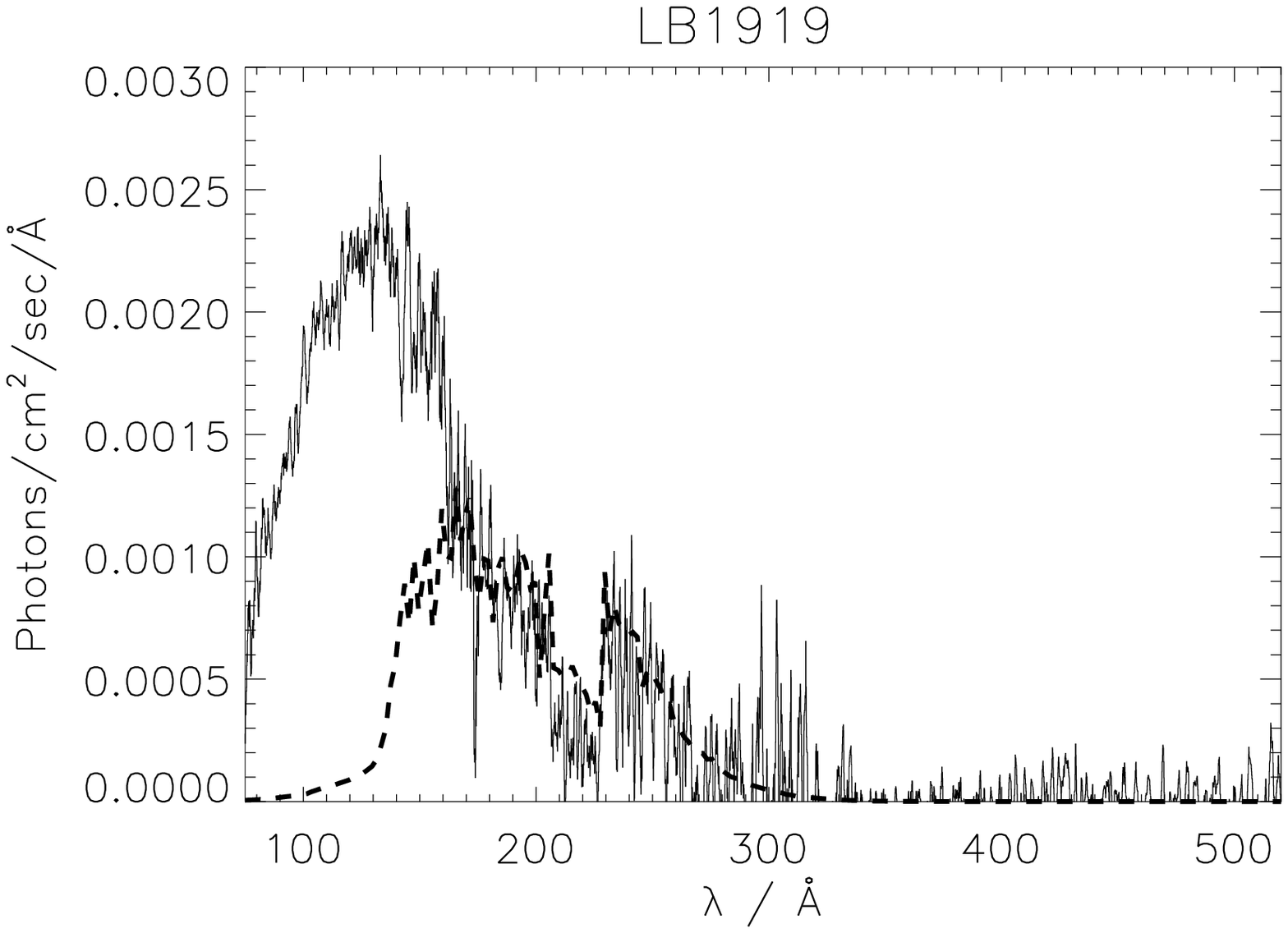}
    \hspace{-3mm}\epsfxsize=0.26\textwidth \epsffile{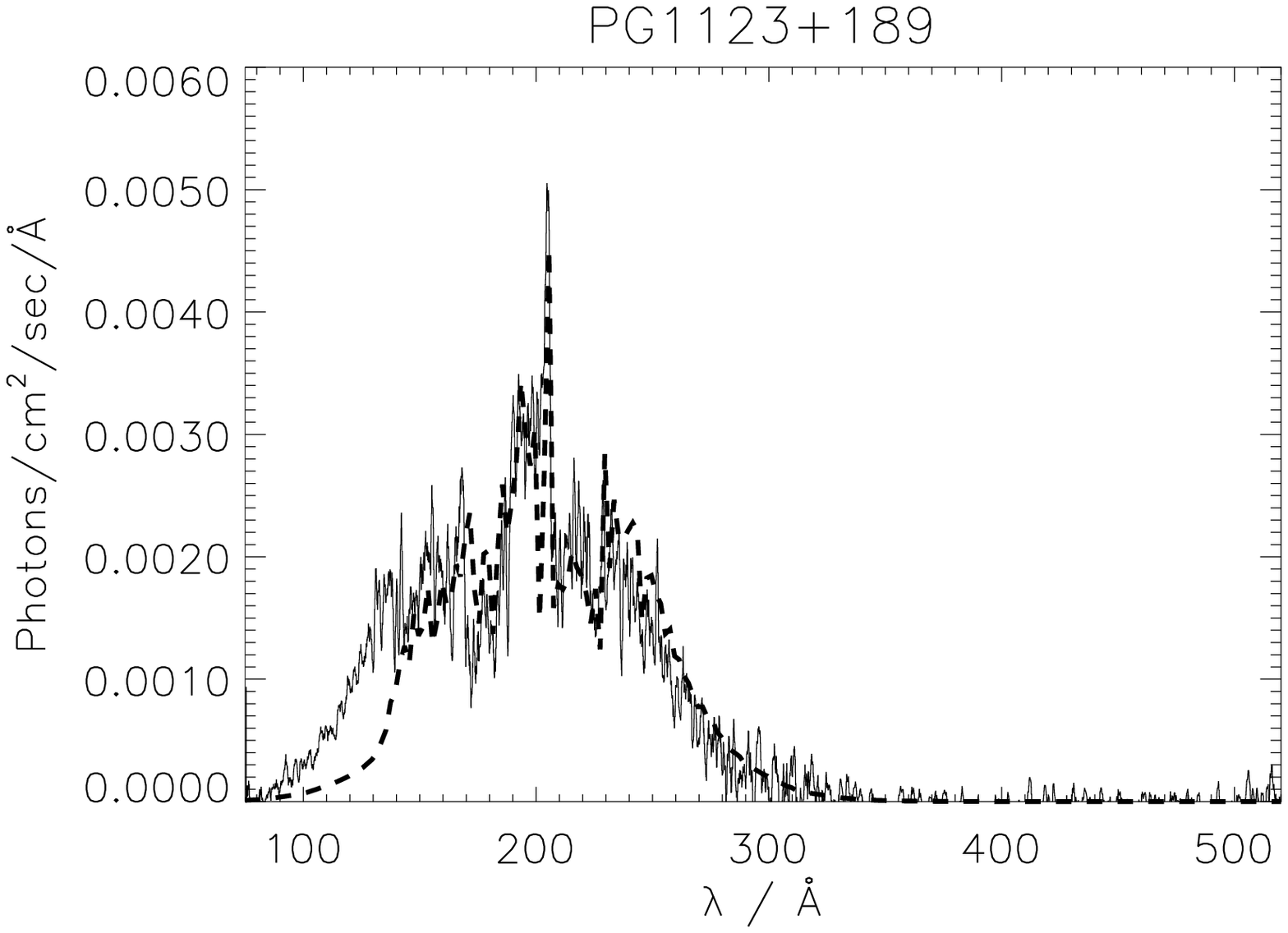}
    }
  \hbox{\hspace{0cm}  
    \epsfxsize=0.26\textwidth \epsffile{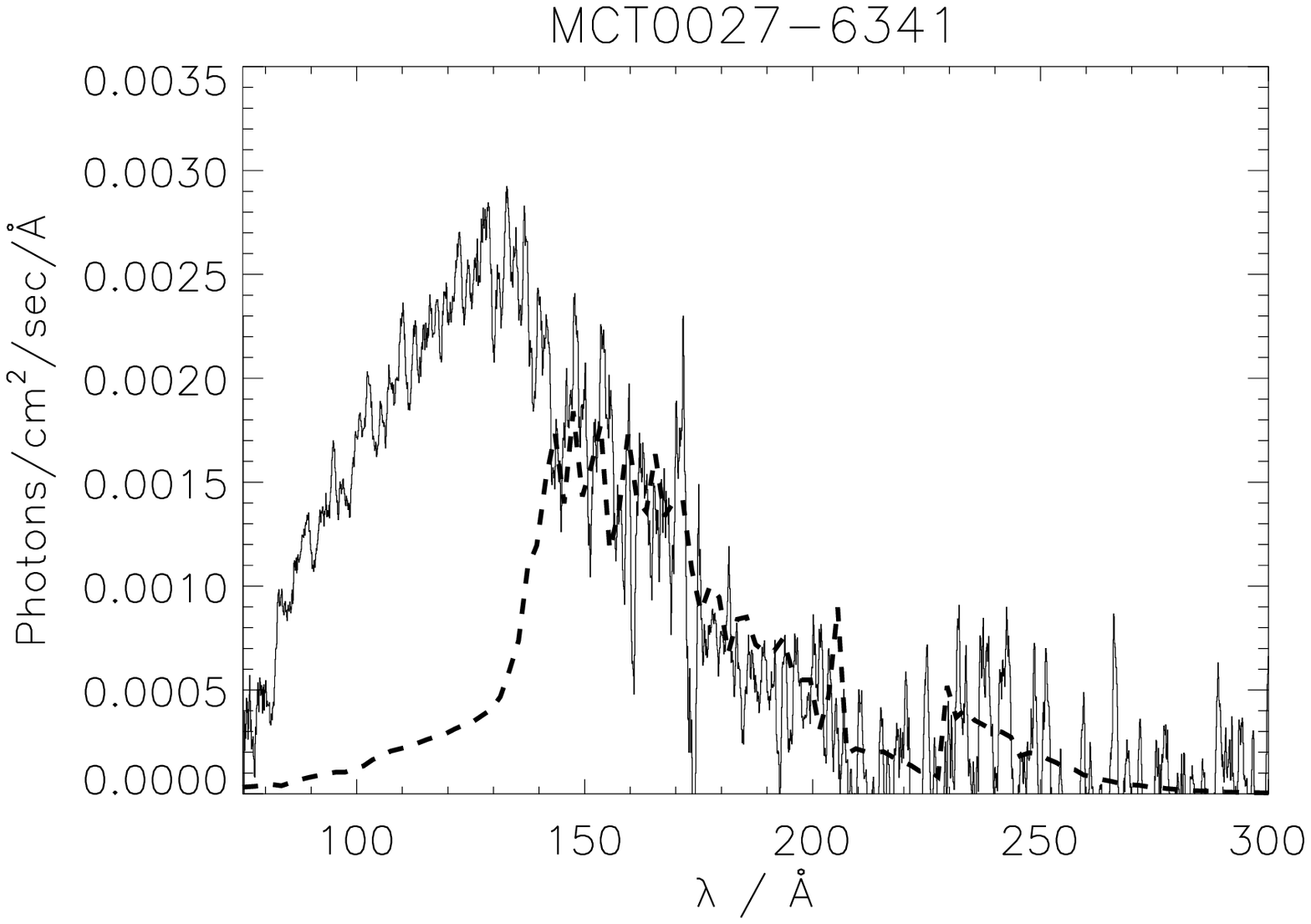}
    \hspace{-3mm}\epsfxsize=0.26\textwidth \epsffile{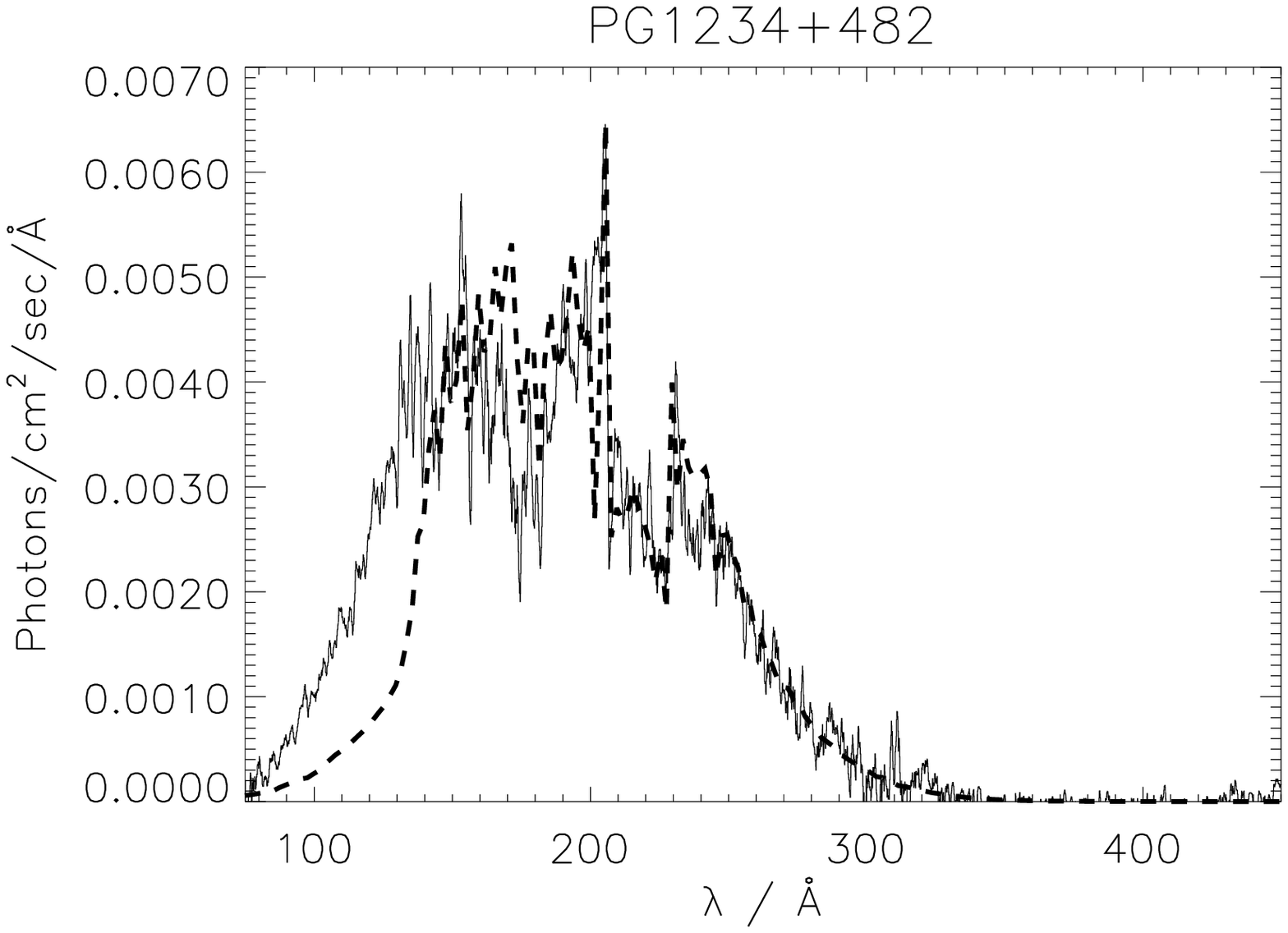}
    \hspace{-3mm}\epsfxsize=0.26\textwidth \epsffile{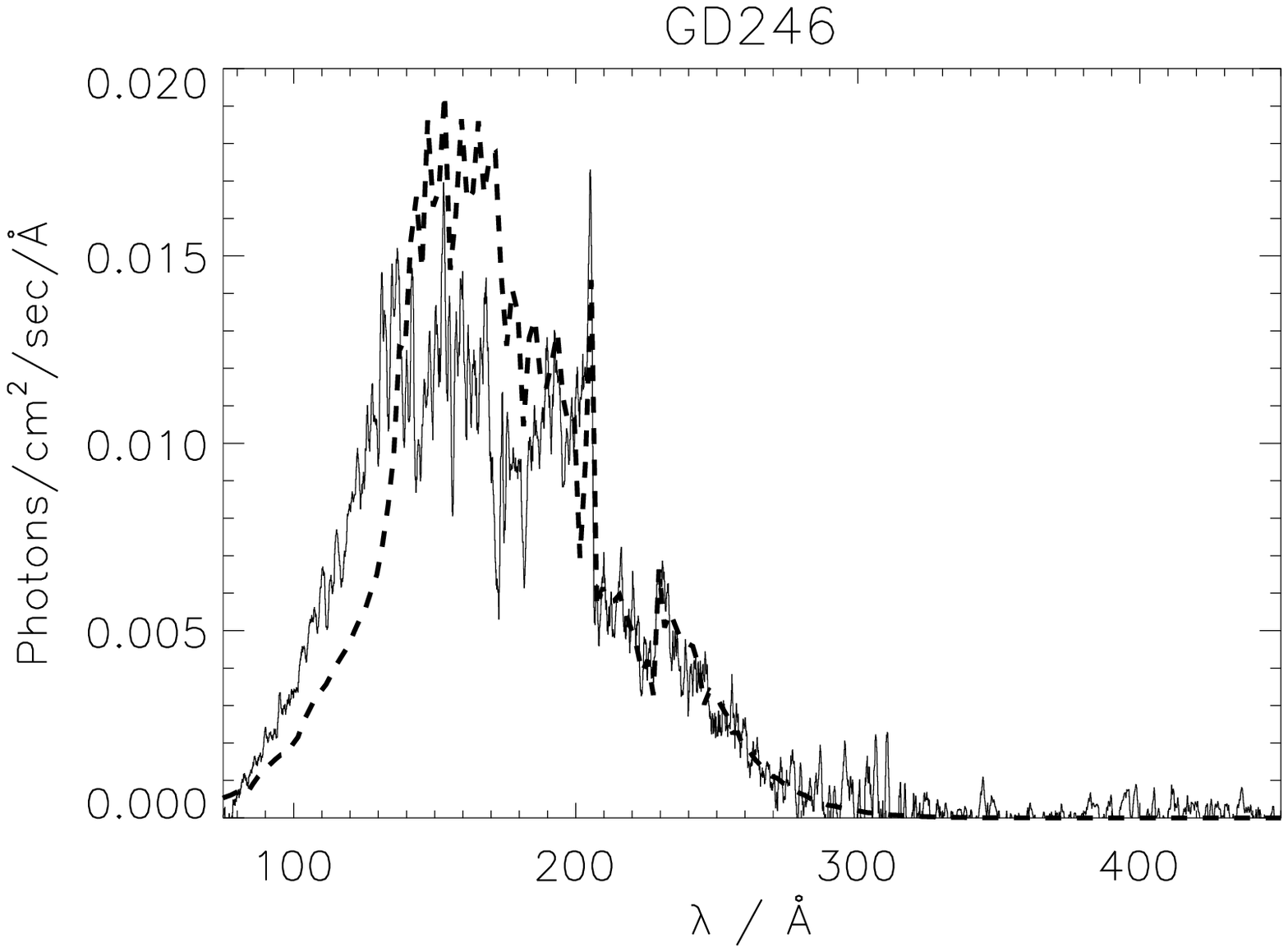}
    \hspace{-3mm}\epsfxsize=0.26\textwidth \epsffile{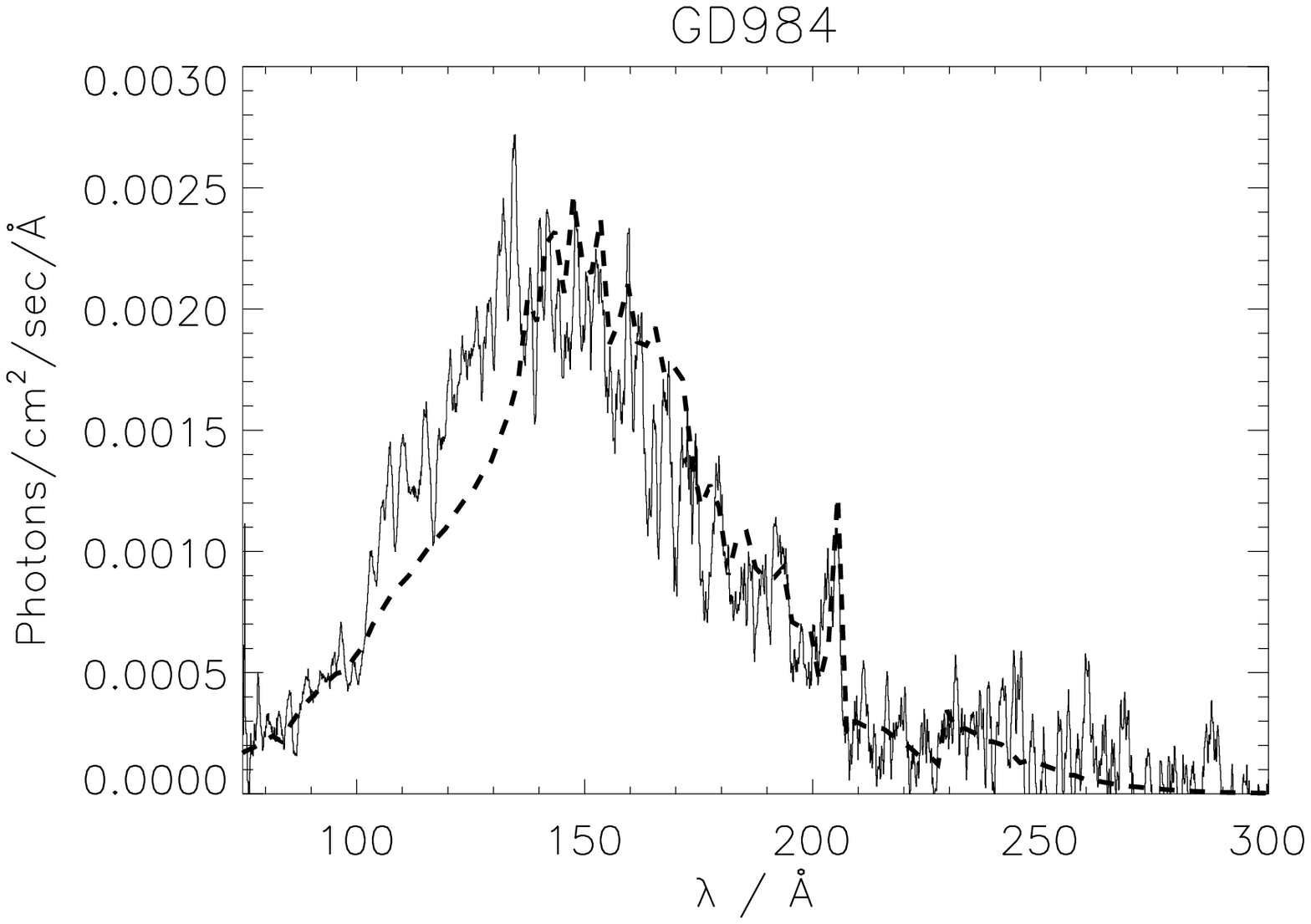}
    }
  \hbox{\hspace{0cm}  
    \epsfxsize=0.26\textwidth \epsffile{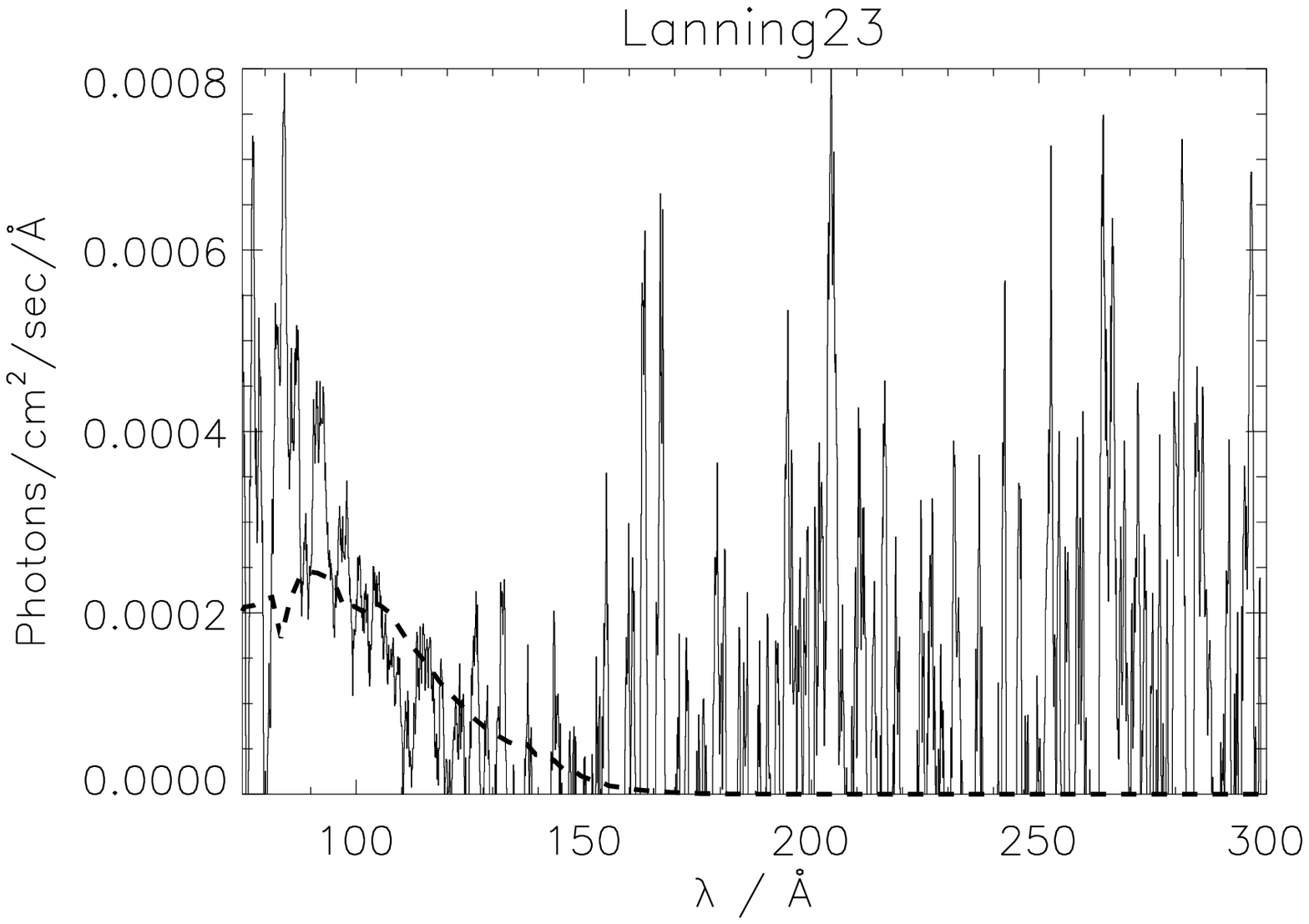}
    \hspace{-3mm}\epsfxsize=0.26\textwidth \epsffile{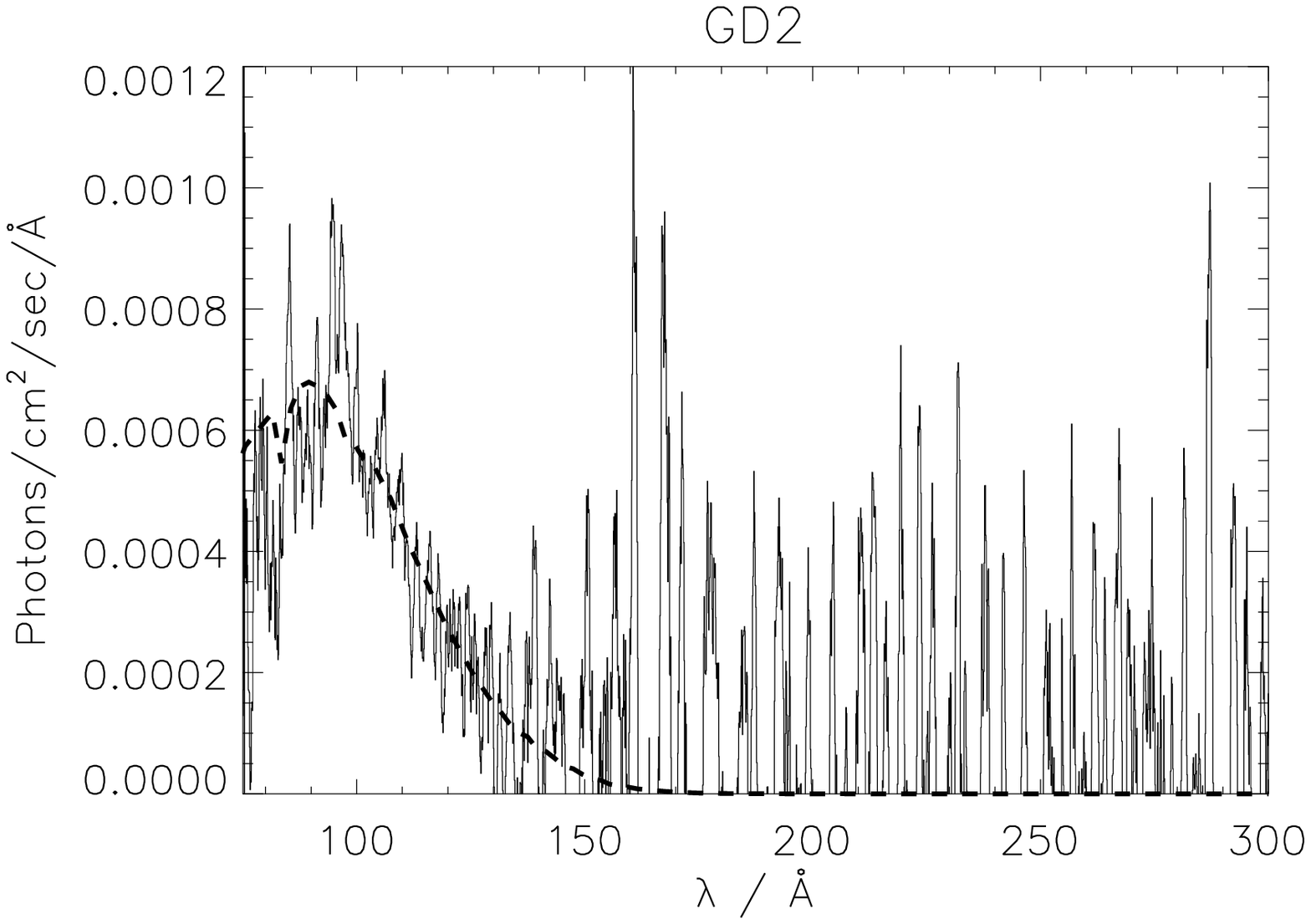}
    \hspace{-3mm}\epsfxsize=0.26\textwidth \epsffile{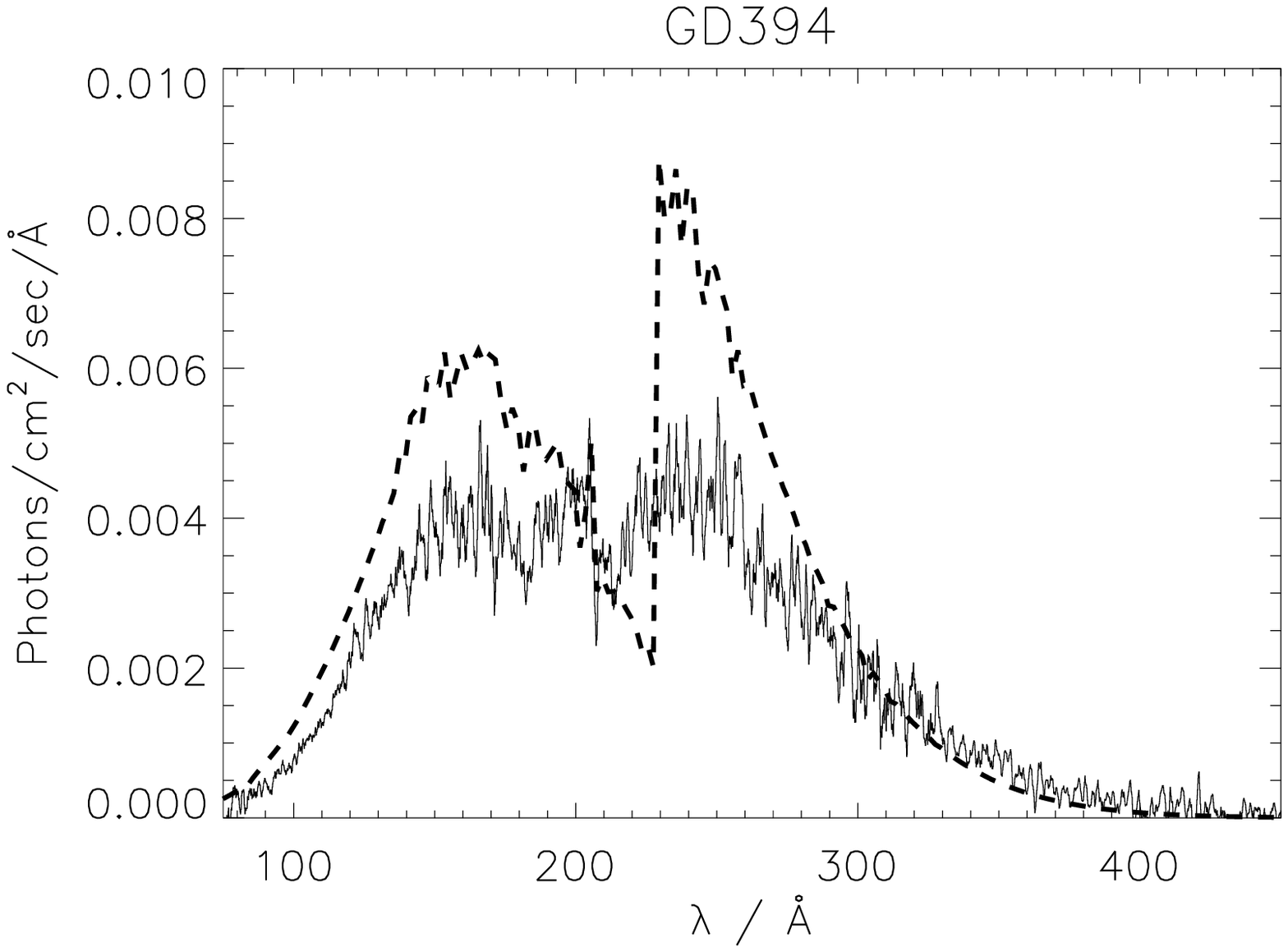}
    \hspace{-3mm}\epsfxsize=0.26\textwidth \epsffile{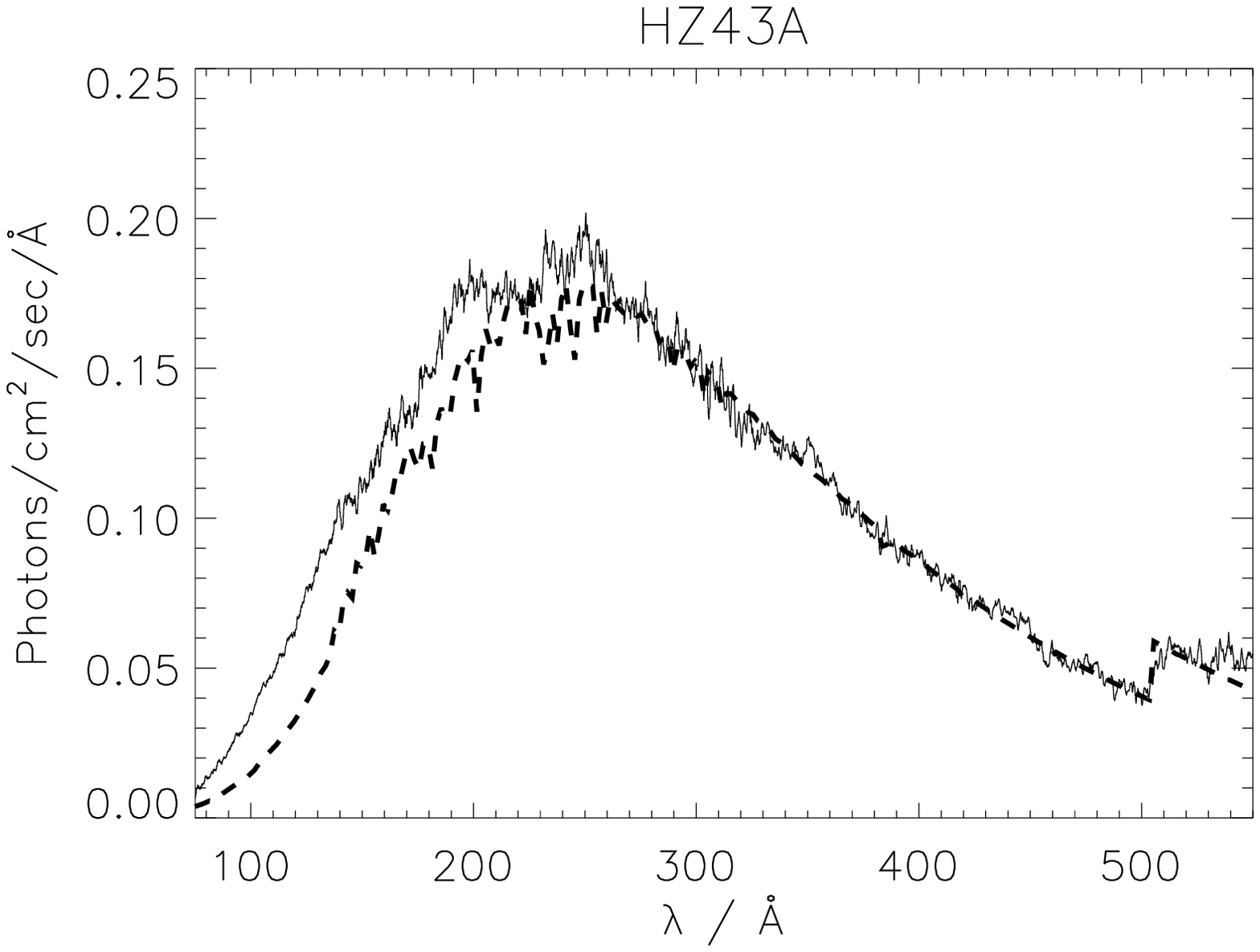}
    }
  \hbox{\hspace{0cm}  
    \epsfxsize=0.26\textwidth \epsffile{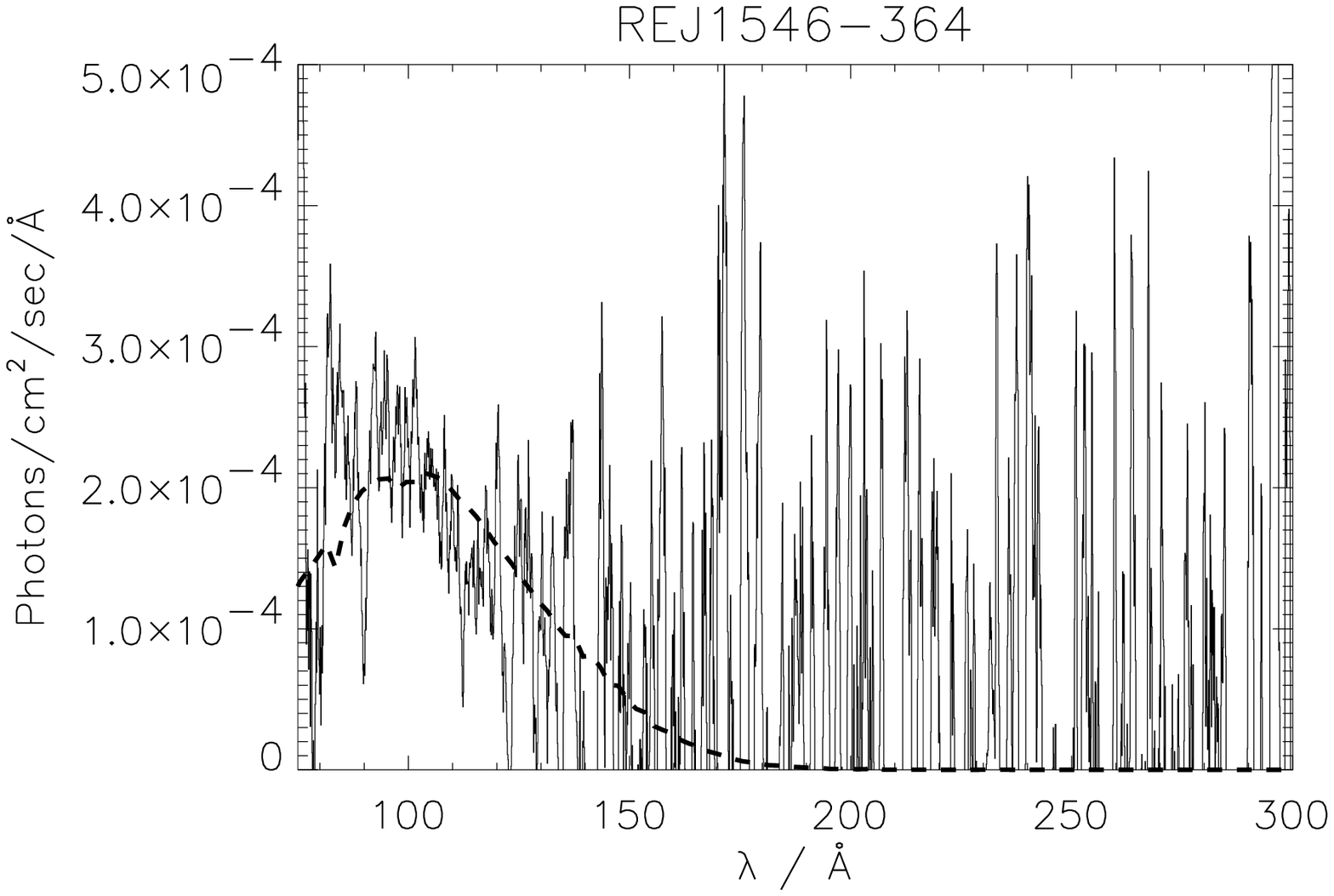}
    \hspace{-3mm}\epsfxsize=0.26\textwidth \epsffile{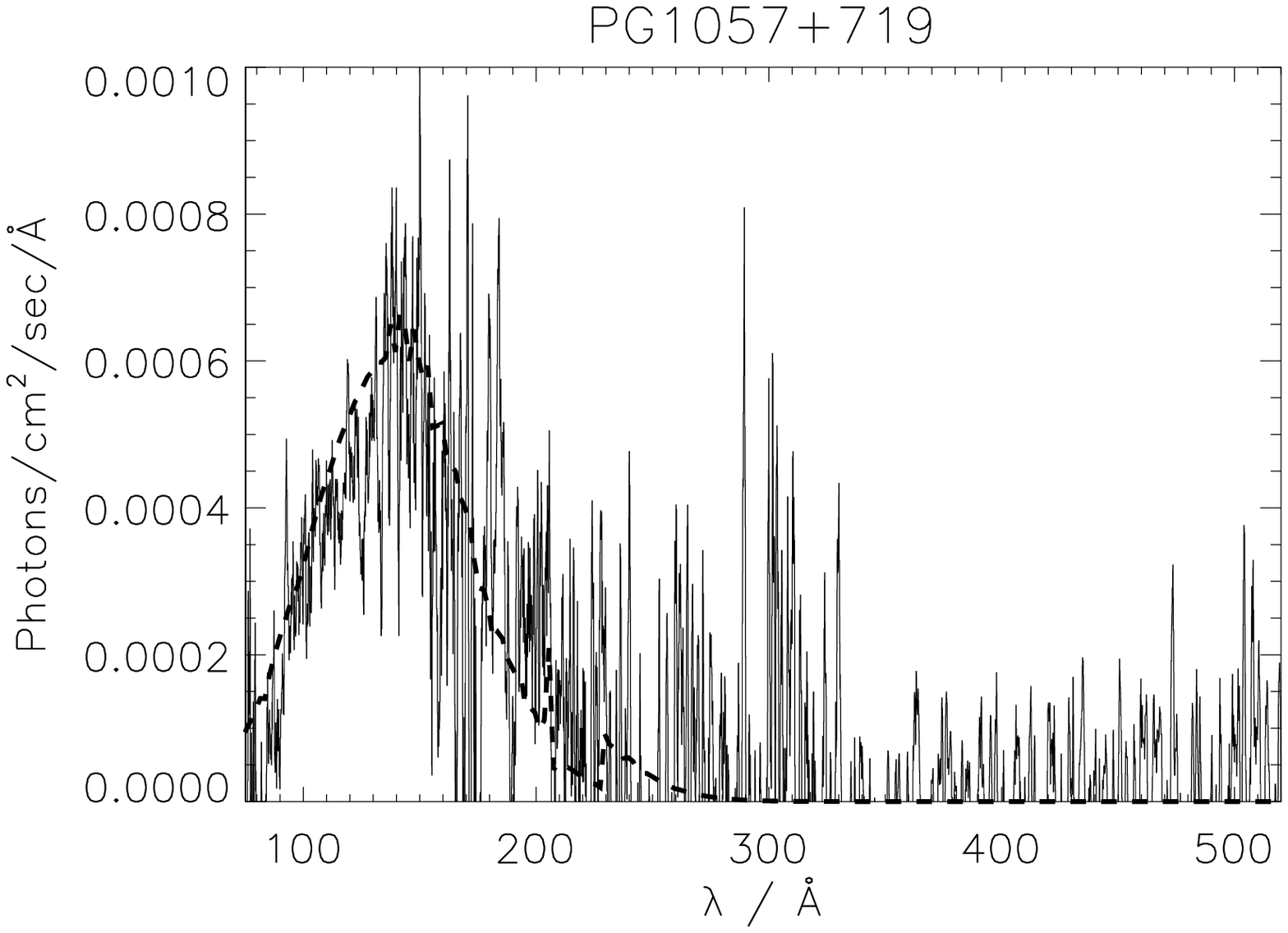}
    \hspace{-3mm}\epsfxsize=0.26\textwidth \epsffile{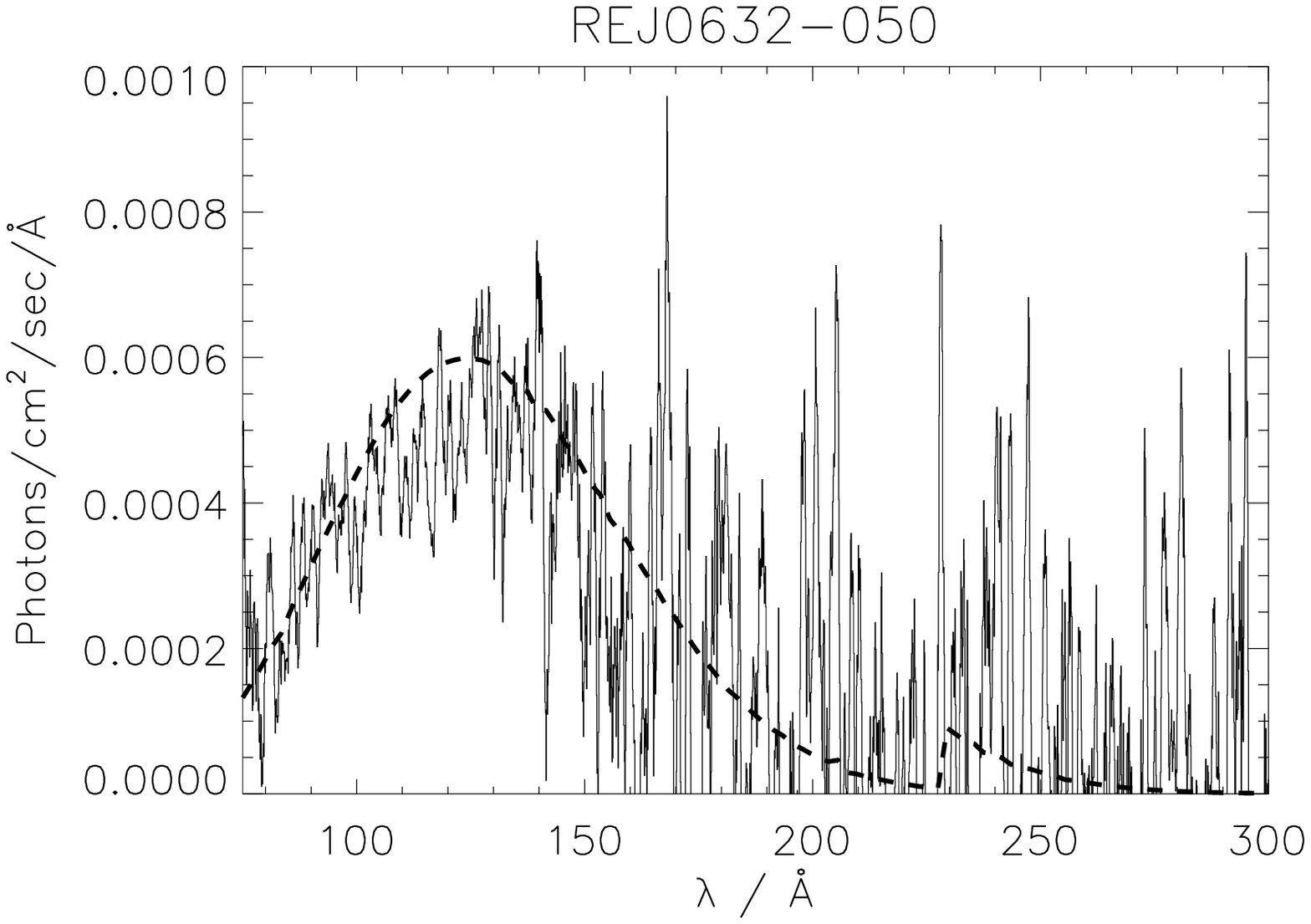}
    \hspace{-3mm}\epsfxsize=0.26\textwidth \epsffile{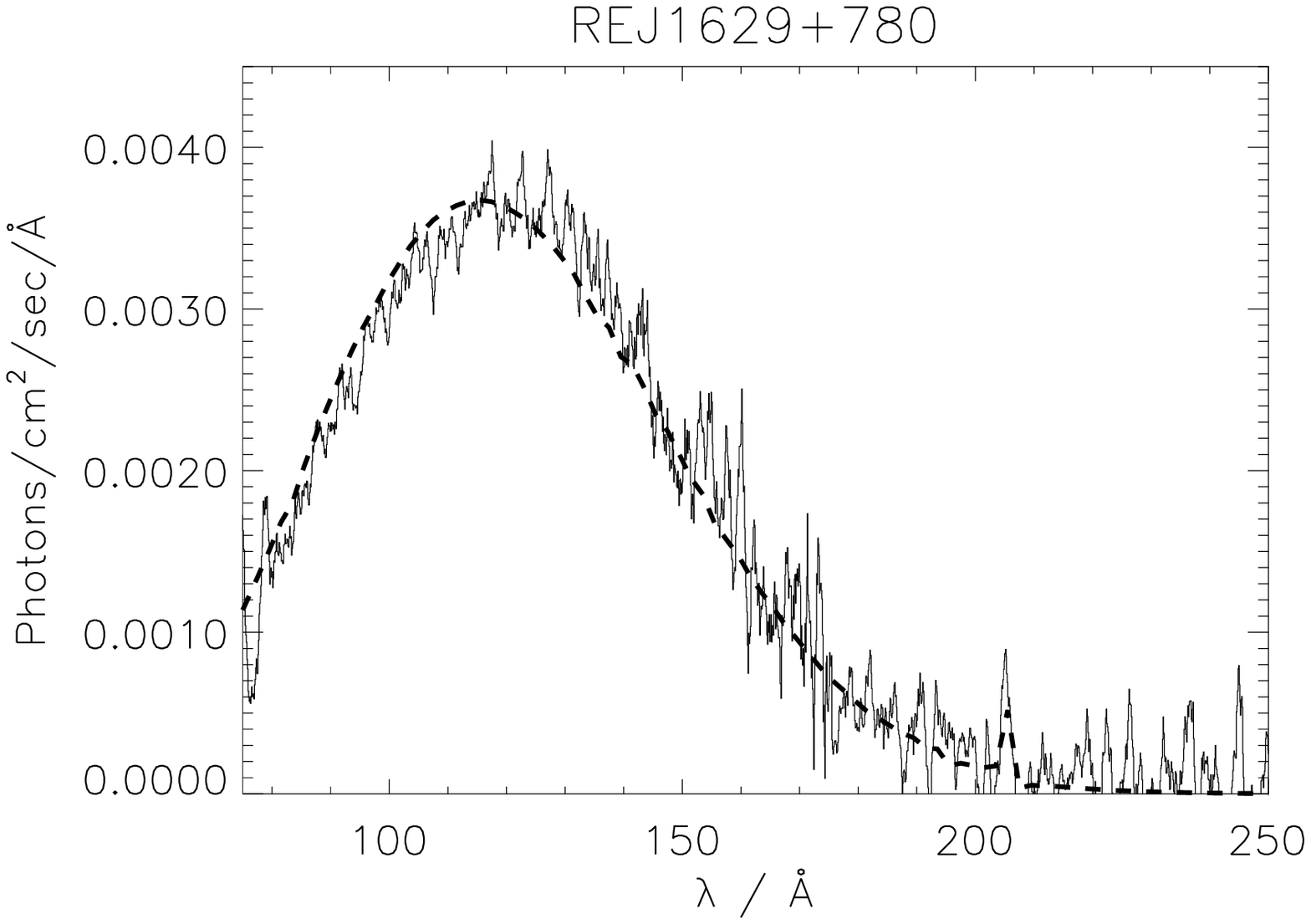}
    }
  \hbox{\hspace{0cm}  
    \epsfxsize=0.26\textwidth \epsffile{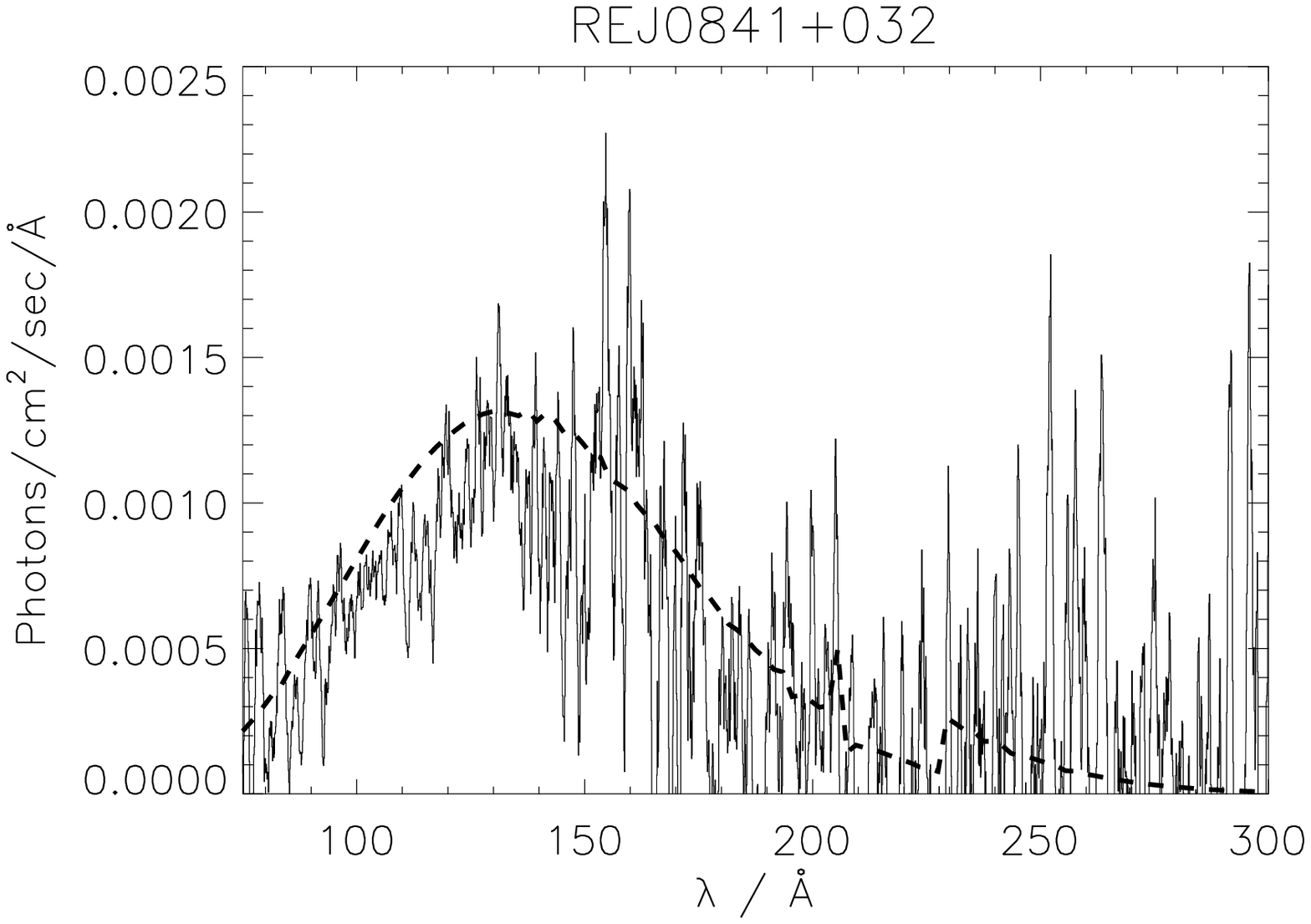}
    \hspace{-3mm}\epsfxsize=0.26\textwidth \epsffile{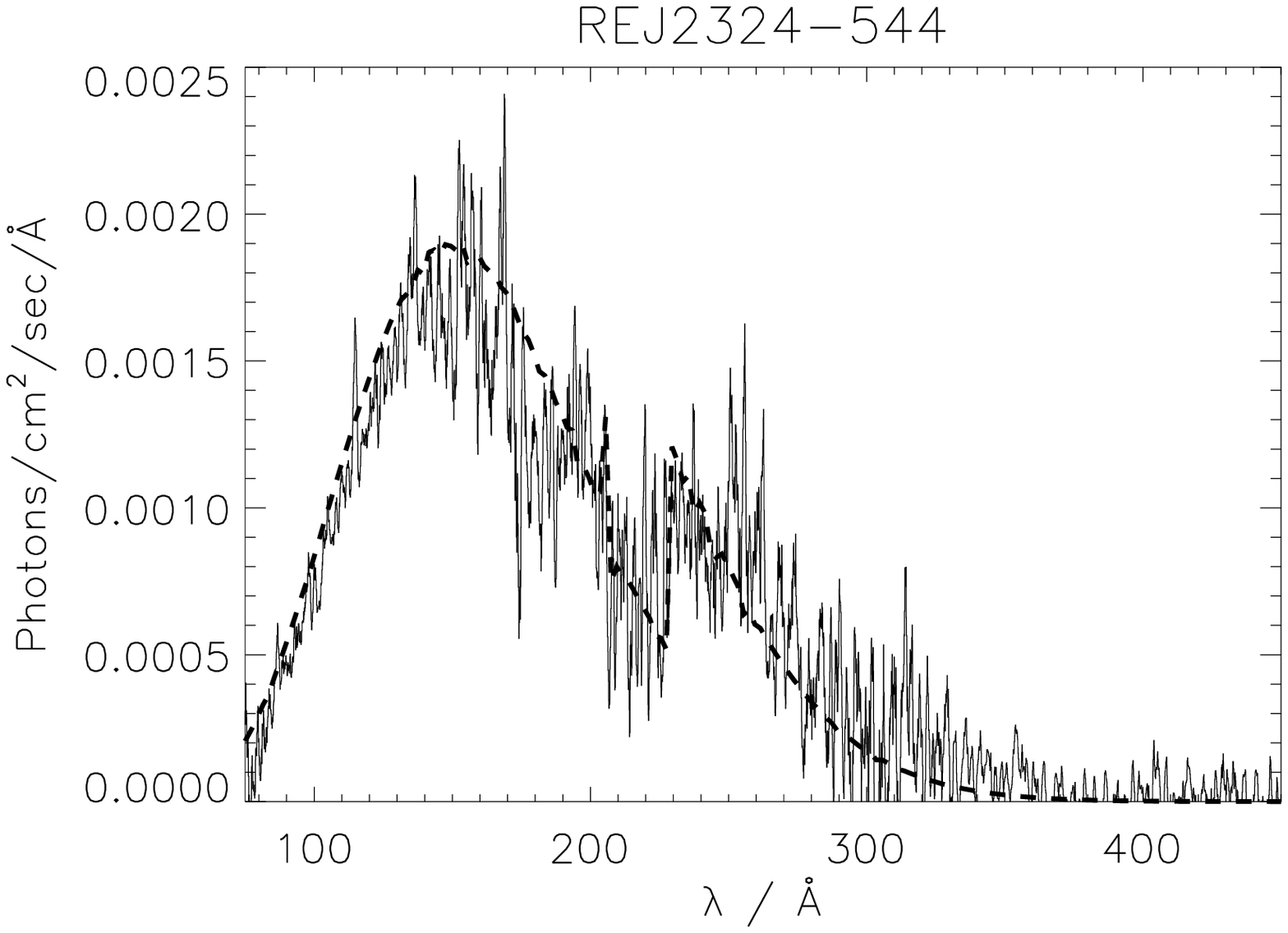}
    \hspace{-3mm}\epsfxsize=0.26\textwidth \epsffile{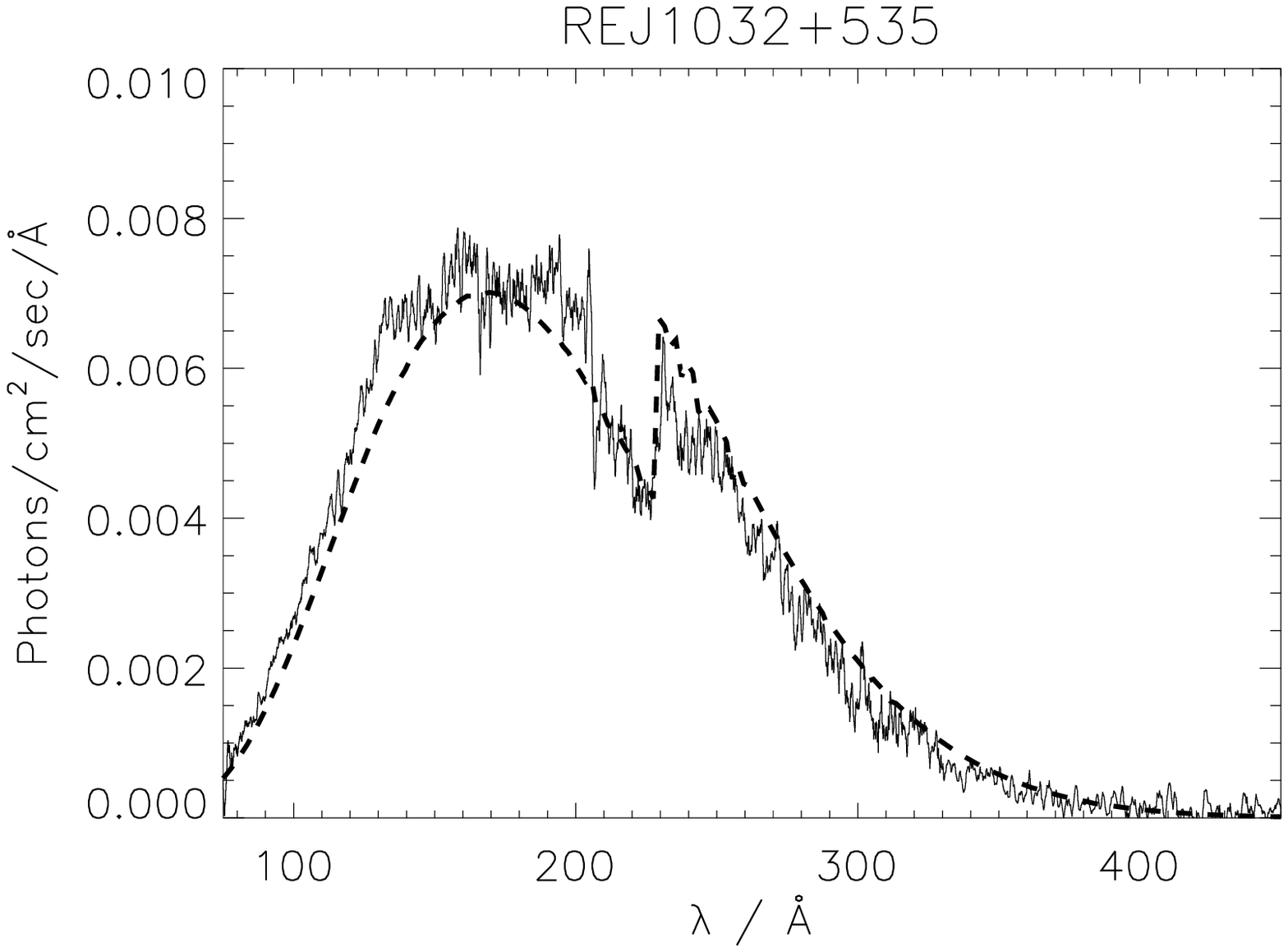}
    \hspace{-3mm}\epsfxsize=0.26\textwidth \epsffile{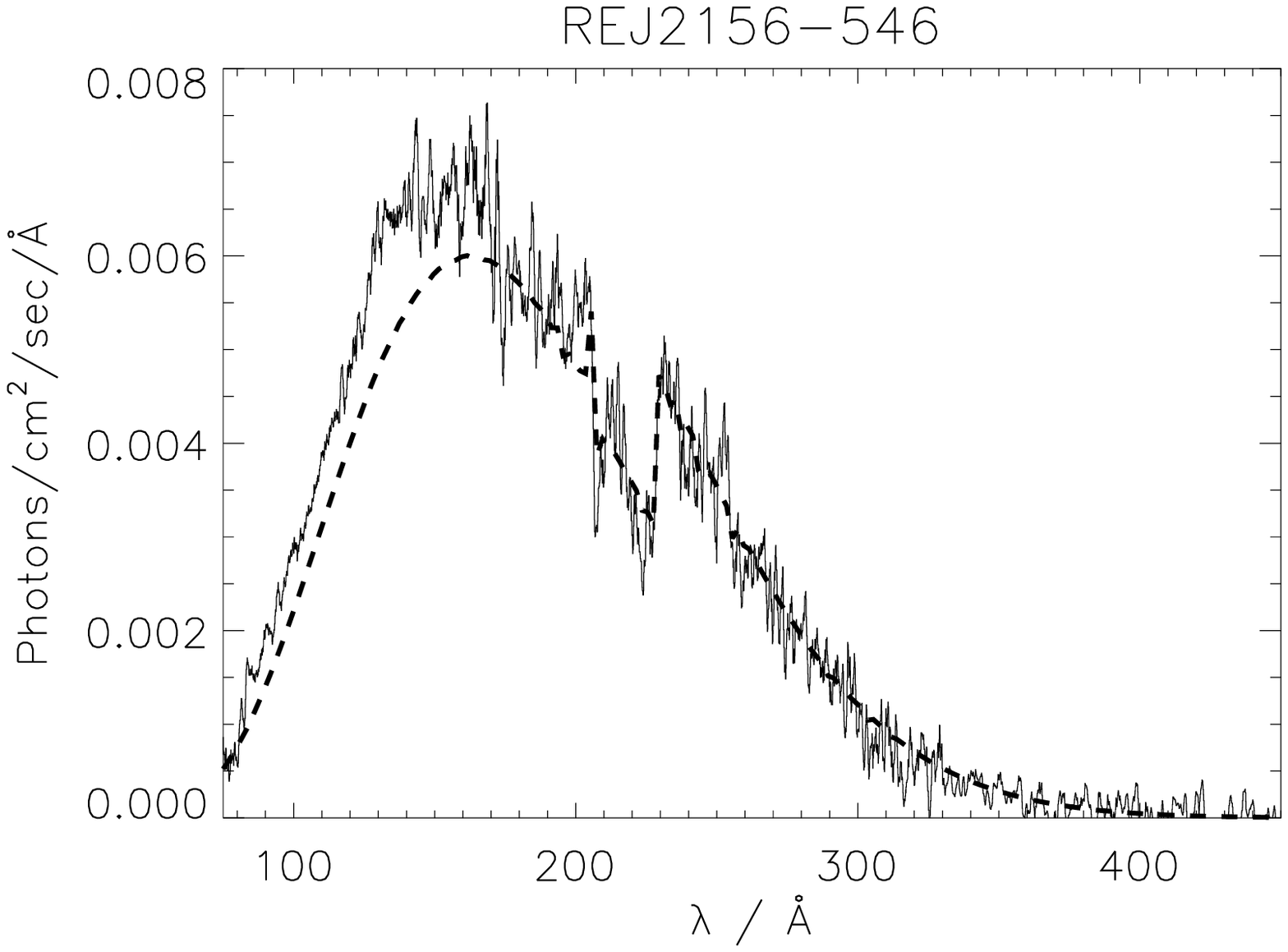}
    }
  \hbox{\hspace{0cm}  
    \epsfxsize=0.26\textwidth \epsffile{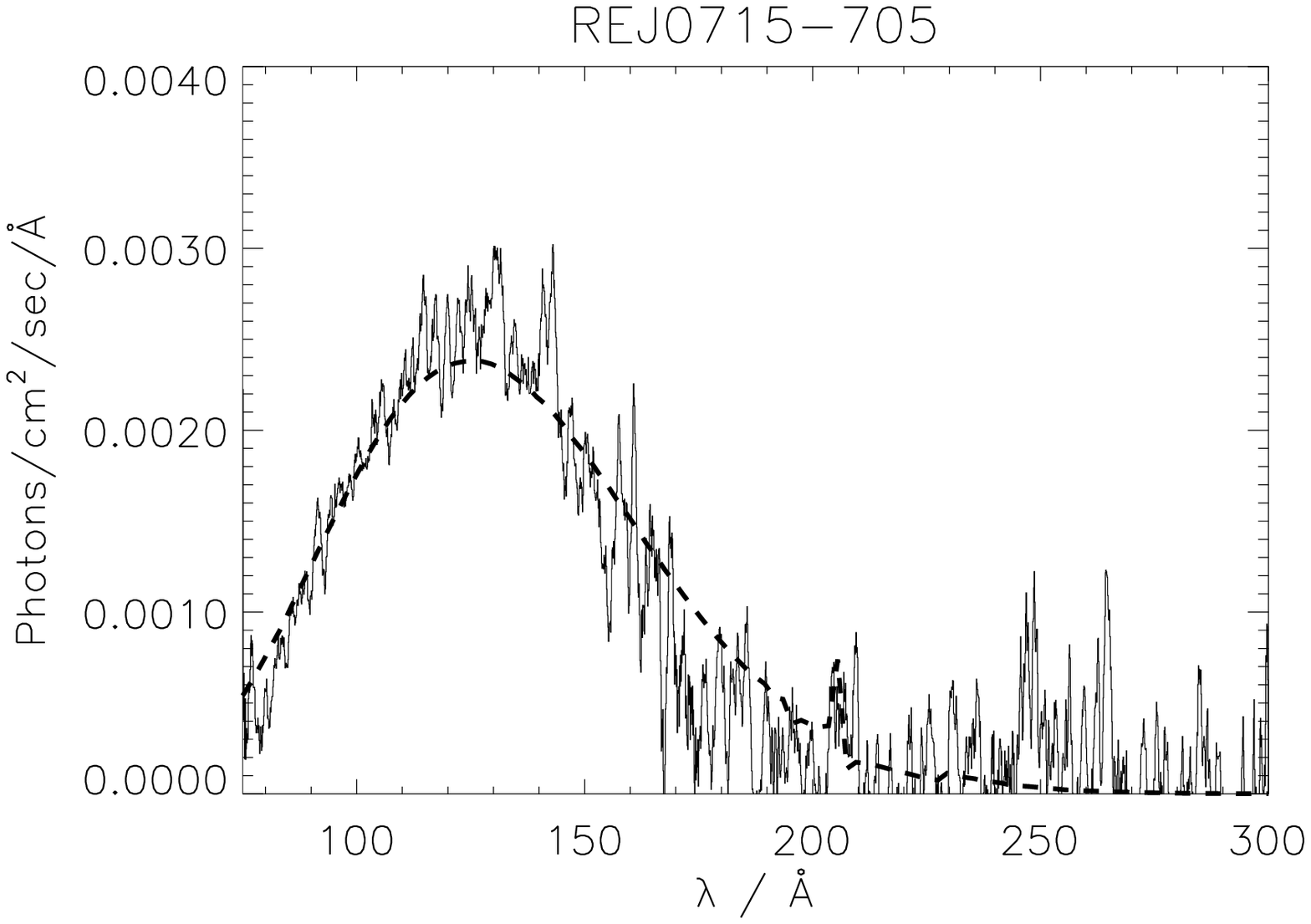}
    \hspace{-3mm}\epsfxsize=0.26\textwidth \epsffile{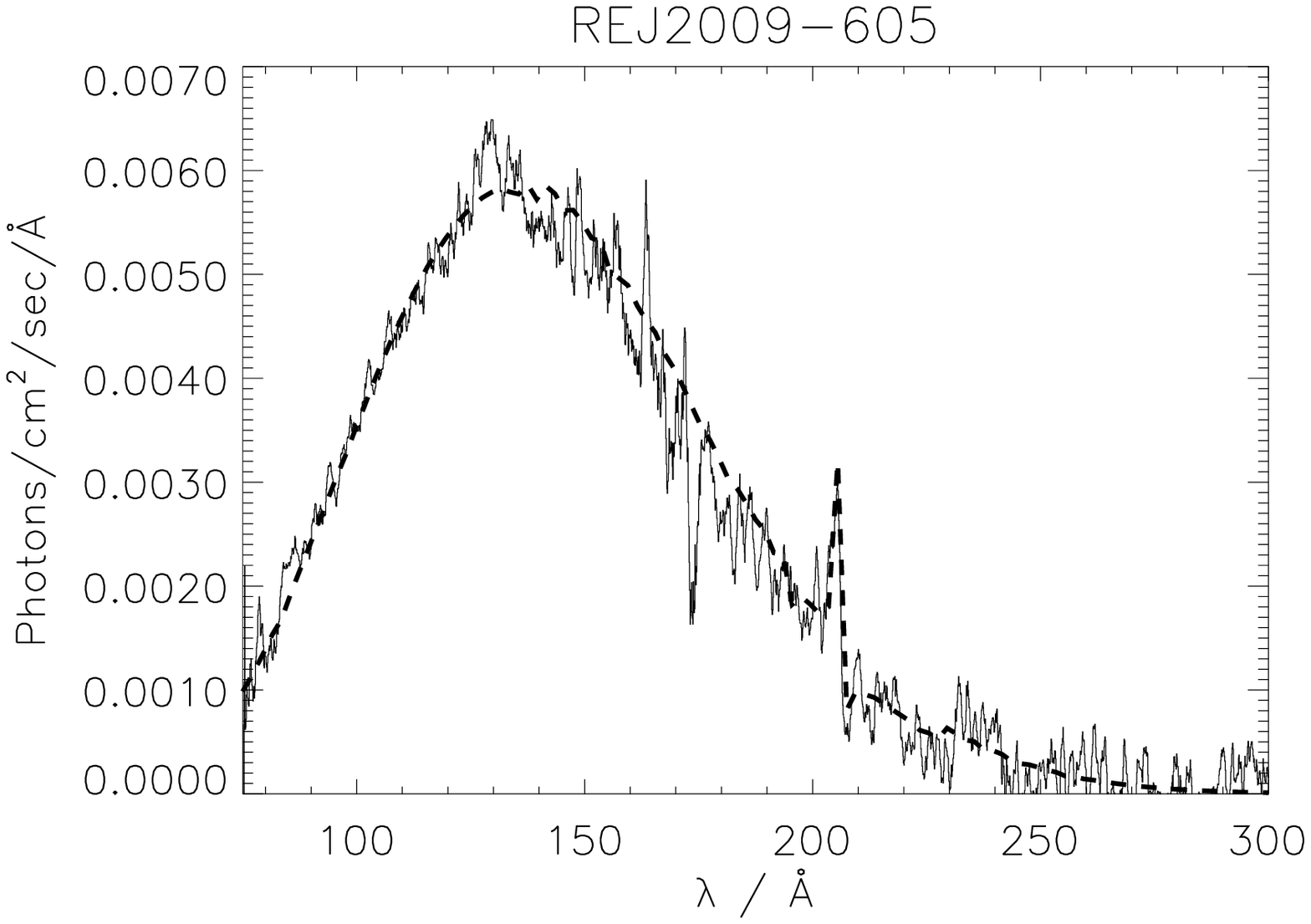}
    }
\end{figure*}

\section{Model matching: Testing the new models and re-analyzing
  the DA sample}
\label{sec:matching}

To directly compare a theoretical flux distribution with the observed one,
it is first re-binned in wavelength and normalized to the observed visual
magnitude. 
The values for m$_{\rm V}$ have been chosen using the compilation of
\citet{mccook:99} and are listed with their individual references in
Table\,\ref{tab:results}.
Then the effect of interstellar absorption is
calculated as sketched in Sect.~\ref{subsec:ism}. 

\subsection{Comparison of theoretical and observed spectra}
\label{subsec:comparison}
The model matching was performed on the mentioned grid which does not in
all parameter ranges correspond to a tight mesh, and no interpolation was
used.  The quality of an individual fit was evaluated by eye.
\par
Of course, changing the interstellar \element{He} column densities
seemingly has a big impact on the overall run of the spectrum. However, the
region most sensitive to metal abundances lies at least partly beyond the
\ion{He}{ii} absorption edge, so once this and the lower energy edges are
fitted all remaining deviations have to be due to either a different flux
level as caused by the effective temperature or a different amount of
absorber content as regulated by the surface gravity in the equilibrium
models.
\begin{table*}[ht]
\caption{New and previous results (\citealt{wolff:98a}) of the EUV
  analysis ordered by decreasing metal index $mi$. N$\rm _H$ is given in 
  $10^{18}$cm$^{-2}$. See Section\,\ref{subsec:metalindex} for an
  explanation of the theoretical metal index $mi$. \vspace{1mm}} 
\label{tab:results}
\begin{tabular}{lll@{}rllllllllll}
\hline
WD--N$^o$ &Name &m$_{\rm V}$     &         
                                 &      &$\log\,g$ &T$\rm _{eff}$
                                 &$mi$&N$\rm _H$&HeI/H&HeII/H&$\log\,g$ &T$\rm _{eff}$
                                 &$m$ \\
          &     &                 &&\multicolumn{7}{c}{new results}
                                 &\multicolumn{3}{c}{previous results} \\ 
\hline
\object{WD\,0621$-$376} &\object{RE\,J0623$-$377}&  12.089&$^1$
                                 &      &7.2  &61000 &3.49  &5.0  &
                                  0.09  &0.15 &7.27&58000&2.0\\
\object{WD\,2211$-$495} &\object{RE\,J2214$-$493}&  11.7&$^1$
                                 &      &7.4  &66000 &3.02  &5.8  &
                                  0.07  &0.15 &7.38&66000&4.0\\
\object{WD\,0232$+$035} &\object{Feige\,24}&  12.56&$^2$
                                 &      &7.6  &59000
                                 &1.22  &2.72  & 
                                  0.068 &0.25  &7.17&58000&1.0\\
\object{WD\,0455$-$282} &\object{MCT\,0455$-$2812}&  13.95&$^3$
                                 &      &7.8  &66000 &1.20  &1.3  &
                                  0.063 &0.3  &7.77&66000&1.0\\
\object{WD\,0501$+$527} &\object{G\,191--B2B}&  11.79&$^2$
                                 &      &7.6  &56000 &0.99  &2.05  &
                                  0.071   &0.3  &7.59&56000&1.0\\
\object{WD\,2331$-$475} &\object{MCT\,2331$-$4731}& 13.1&$^4$ 
                                 &      &7.6  &56000 &0.99  &8.5  &
                                  0.08  &0.1  &8.07&56000&0.75-1.\\
\object{WD\,1056$+$516} &\object{LB\,1919}&  16.8&$^4$
                                 &      &8.2 &70000 &0.61  &16.0& 
                                  0.04  &0.05 &&69000&0.1\\
\object{WD\,1123$+$189} &\object{PG\,1123$+$189}&14.13&$^1$  
                                 &      &7.9  &54000 &0.43  &11.9  &
                                  0.09  &0.052 &7.63&54000&0.4\\
\object{WD\,0027$-$636} &\object{MCT\,0027$-$6341}&  15.0&$^4$
                                 &      &8.2 &64000 &0.42  &21.5 &
                                  0.068  &0.06 &7.96&64000&0.2\\
\object{WD\,1234$+$482} &\object{PG\,1234$+$482}&  14.38&$^5$
                                 &      &8.1  &56000 &0.31  &11.7  &
                                  0.09  &0.05 &7.67&56000&0.2\\
\object{WD\,2309$+$105} &\object{GD\,246}&  13.09&$^2$
                                 &      &8.2  &56000 &0.25 &18.0  &
                                  0.05   &0.03  &7.81&59000&0.25\\
\object{WD\,0131$-$164} &\object{GD\,984}&  13.98&$^6$
                                 &      &8.1  &50000 &0.20  &22.0  &
                                  0.068 &0.035 &7.67&50000&0.2\\
\object{WD\,2247$+$583} &\object{Lanning\,23}&  14.26&$^7$
                                 &      &8.3  &56000 &0.20  &40.0  &
                                  0.068   &0.03  &7.84&59000&0.25\\
\object{WD\,0004$+$330} &\object{GD\,2}&  13.85&$^2$
                                 &      &8.25  &50000 &0.14  &82.4  &
                                  0.068 &0.01 &7.63&49000&$<0.1$\\
\object{WD\,2111$+$498} &\object{GD\,394}&  13.09&$^8$
                                 &      &7.9  &40000 &0.13  &6.5  &
                                  0.07  &0.15 &7.94&39600&0.25\\
\object{WD\,1314$+$293} &\object{HZ\,43A}&  12.914&$^9$
                                 &      &8.3  &50000 &0.13  &0.9  &
                                  0.06 &0.0&7.99&50800&0.0\\
\object{WD\,1543$-$366} &\object{RE\,J1546$-$364}&15.81&$^{10}$
                                 &      &8.3  &50000 &0.13  &50.4  &
                                  0.068   &0.03 &8.88&45200&0.0\\
\object{WD\,1057$+$719} &\object{PG\,1057$+$719}&  14.68&$^1$
                                 &      &8.0  &42000 &0.12  &20.7  &
                                  0.068  &0.052  &7.90&41500&0.1\\
\object{WD\,0630$-$050} &\object{RE\,J0632$-$050}&  15.537&$^1$
                                 &      &8.2  &44000 &0.09  &30.1  &
                                  0.01  &0.055&8.39&44100&0.0\\
\object{WD\,1631$+$781} &\object{RE\,J1629$+$780}&13.03&$^{11}$
                                 &      &8.15  &42000 &0.09  &35.0  &
                                  0.068 &0.0 &7.79&44600&0.05\\
\object{WD\,0838$+$035} &\object{RE\,J0841$+$032}&  14.48&$^1$
                                 &      &8.1  &40000 &0.08 &16.0  &
                                  0.068  &0.052&7.78&38400&0.0\\
\object{WD\,2321$-$549} &\object{RE\,J2324$-$544}&  15.2&$^1$
                                 &      &8.2  &42000 &0.08  &9.5  &
                                  0.05  &0.06 &7.94&45000&0.05\\
\object{WD\,1029$+$537} &\object{RE\,J1032$+$535}&  14.455&$^1$
                                 &      &8.3  &44000 &0.08        &7.5  &
                                  0.005  &0.04 &7.77&44000&0.0\\
\object{WD\,2152$-$548} &\object{RE\,J2156$-$546}&  14.44&$^1$
                                 &      &8.3  &44000 &0.08  &7.0  &
                                  0.04 &0.04 &7.91&44000&0.0\\
\object{WD\,0715$-$703} &\object{RE\,J0715$-$705}&  14.178&$^1$
                                 &      &8.3  &44000 &0.08  &21.9  &
                                  0.068  &0.015 &8.05&44000&0.0\\
\object{WD\,2004$-$605} &\object{RE\,J2009$-$605}&  13.6&$^1$
                                 &      &8.3 &42000 &0.06  &17.5  &
                                  0.06  &0.015 &8.16&41900&0.0\\
\hline\hline
\end{tabular}
$^{1}$~\citet{marsh:97a}
$^{2}$~\citet{kidder:91}
$^{3}$~\citet{barstow:94}
$^{4}$~\citet{pounds:93}
$^{5}$~\citet{green:80}
$^{6}$~\citet{wesemael:95}
$^{7}$~\citet{vennes:97a}
$^{8}$~\citet{bergeron:92}
$^{9}$~\citet{bohlin:95}
$^{10}$~\citet{vennes:96b}
$^{11}$~\citet{schwartz:95}
\end{table*}
\subsection{Revised atmospheric parameters}
\label{subsec:revisedparameters}
The present model matching led to new atmospheric parameters where mainly
the surface gravities differ from former specifications. The improvement of
many of the individual fits (see Fig.\,\ref{fig:fits}) as compared to fits
with homogeneous models \citep[in][]{wolff:98a,wolff:phd,dreizler:99a}
vindicates these adjustments.
\par
With homogeneous models, discrepancies were encountered especially for
higher metallicity stars in the shorter wavelength regimes, which could not
be adjusted by a different metallicity value as this would have deteriorated
the quality of the fit at other wavelengths. This difficulty is resolved
with the new models, as they offer different abundances at different
formation depths. Consequently they can often better reproduce especially
these formerly delicate regions. 
Fig.\,\ref{fig:mct2331} shows an example. 
\par
Fig.\,\ref{fig:fits} displays all observations
with the best-fit model spectra over-plotted, both shown at 1\AA\
resolution. Table\,\ref{tab:results} lists the parameters of the best-fit
model for each object, along with the interstellar column densities used to
produce the plots. 
These are to be regarded as preliminary as fitting improved
models to the spectra may yet again yield different values. 
High \ion{He}{ii} column densities, found 
throughout the sample, may be an indication that some opacities are
still missing.
It additionally lists a quantity denoted $mi$, which stands for
the \emph{metal~index} that is to be introduced now.
\clearpage
\begin{figure*}[th]
  \epsfxsize=13cm \epsfbox{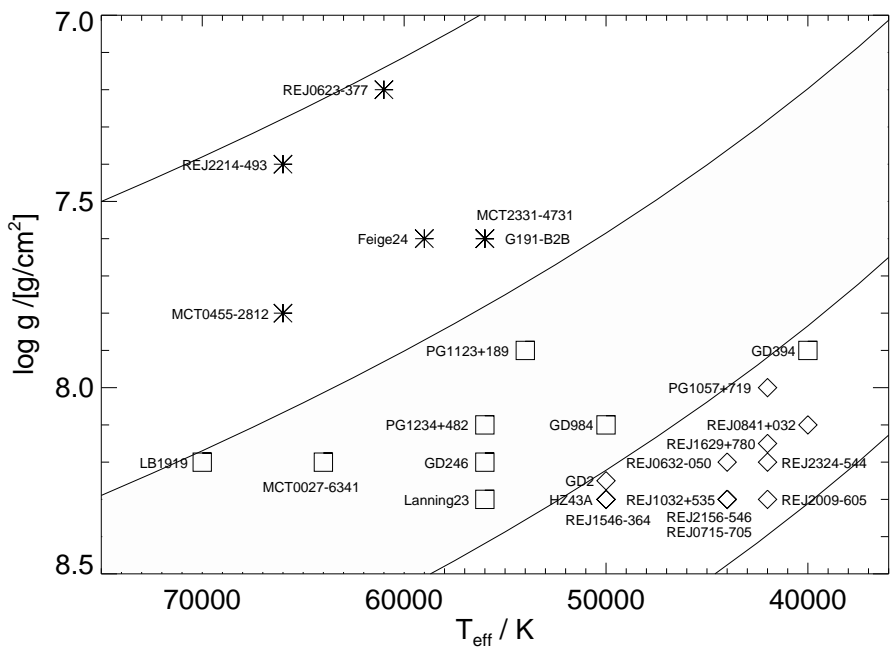}
  \caption[]{Revised parameters of the program
  stars in the [$T_{\rm eff}$,$\log{g}$] plane. Lines indicate constant
  values of the metal index  as defined in
  Sect.\,\ref{subsec:metalindex}, $mi=4., 0.65, 0.15, 0.05$,
  respectively. The shaded area indicates the 
  metallicity of DA white dwarfs of the GD\,246 group of
  \citet{wolff:98a}. These as well as the
  objects classified as irregular fall into this regime and are marked by
  squares. Stars above that region belong to the G\,191--B2B-group
  (identified by starry symbols), stars below to the pure-hydrogen-group 
  (marked by diamond-shaped symbols). See also Sect.\,\ref{subsec:sample}.} 
  \label{fig:metalindex}
\end{figure*}

\subsection{Metal index}
\label{subsec:metalindex}
With depth dependent abundances, it is difficult to quote one
representative value for each element. To evade this problem, we have
introduced a photospheric parameter dependent quantity that relies entirely
on the predictions by diffusion theory. As explained above, the absorber
content of the models is determined from the equilibrium between the
radiative and the gravitational forces.  The radiative acceleration scales
with $T_{\rm eff}^4$ due to the occurrence of the Eddington flux in the
lower expression of Eq. \ref{geffisgradis}, the effective gravitational
acceleration scales with $g$ as is evident from the upper expression in the
same equation. The absorber content should thus be comparable along lines
of constant $T_{\rm eff}^4/g$.  To map this ratio onto a dimensionless and
easier-to-read parameter, we define
\begin{equation}
  \label{defmetalindex}
  mi=4\cdot10^{-12}\cdot T_{\rm eff}^4/g\,\,\,/[{\rm K^4 s^2/cm}]
\end{equation}
where the leading factor has been chosen to yield a result of the order of
one for the photospheric parameters of the standard star G\,191--B2B. Lines
of constant $mi$ are displayed in Fig.\,\ref{fig:metalindex}.
\par
It should be emphasized that this quantity is not derived from the
actual calculated equilibrium abundances in the models. It is based on a
much simpler evaluation of just the same idea that underlies the
construction of the models. Knowing this, it is all the more surprising how
well the metal indices $mi$ at the newly determined parameters match the
metallicities $m$ that \citet{wolff:phd} has derived for the
objects.
%
\section{Discussion}
\label{sec:discussion}

Our new models can reproduce the observed EUV spectra of hot DA white
dwarfs. Since the new models predict the chemical composition from the
equilibrium between sedimentation and radiative acceleration, the number of
free parameters is drastically reduced to the effective temperature and
surface gravity only. The good agreement is a strong evidence that the
interplay of these two processes defines the chemical composition and
stratification. The agreement is, however, not in all cases better than
with chemically homogeneous models, which requires a more detailed
discussion.
\par
Stars similar to G\,191--B2B \citep[the group one from][]{wolff:98a} can be
reproduced significantly better with our 
self-consistent stratified models. This result is not surprising since the
first and successful application of these models to G\,191--B2B
\citep{dreizler:99a} motivated this work in the first place. Our new models
can also reproduce the stars classified as pure hydrogen
\citep[group three from][]{wolff:98a}, demonstrating that the vanishing of trace
amounts of metals is correctly reproduced. Within this group, however, the
surface gravity had to be increased by up to 0.5\,dex in order to achieve a
good fit (see also Table\,\ref{tab:results}).
\par
The fits of PG\,1123$+$189, GD\,246 and GD\,984 are 
not completely satisfying but deviations are of the same order as compared
to fits with homogeneous models. 
The fits for HZ\,43A, RE\,J1032$+$535, RE\,J2156$-$546 and RE\,J2009$-$605
are not quite as good as the ones which could be achieved with spectra
derived from the grid of homogeneous models.
Further improvements of our models 
(see Sect.\,\ref{subsec:perspective})
may resolve these slight discrepancies.
\par
In four cases (LB\,1919, MCT\,0027$-$6341, PG\,1234$+$482, and GD\,394),
using the new models results in a worse fit than before.  
One may be tempted to explain this by the fact that our model grid is not
yet extended enough to find a reasonable agreement. However, significant 
individual deviations between models and observation could also originate
due to physical reasons. As discussed in Sect.\,\ref{subsec:scope}, competing
processes can disturb the equilibrium between radiative acceleration and
gravitational settling. As is evident from the overall good fit
of our models, these are unimportant in general but individual exceptions are
possible. In the case of LB\,1919 for example, an unusually high rotation
rate has been suspected \citep{Finley:1997}. Prior to a final decision, we
will have to implement several improvements to our models.
\par

\subsection{Remarks on $g$ and $g_{rad}$}
\label{subsec:remarks}

As mentioned above, our surface gravities are in several cases higher than
in earlier analyses. Those analyses had evaluated 
the Balmer lines for a determination of the surface gravity. 
Should the surface gravities determined with our models prove to be
reliable, they would open up a new possibility:
The surface gravity governs the gravitational settling. Through the
equilibrium condition, the amount of trace metals in the atmosphere is
determined. In effect, we therefore evaluate the metal abundances in order
to derive the surface gravity.
The new models with stratified metal abundances can modify the Balmer lines via
the changed atmospheric structure. This could explain the deviations in the
surface gravity determination. A more detailed investigation is in progress.
Our proposed method would be
a very sensitive indicator but it is, unfortunately, also
sensitive to systematic errors in the calculation of the radiative acceleration. 
The shifts in $\log\,g$ necessary to obtain good fits for several of the
objects consequently could indicate that the accuracy of the $g_{rad}$
calculation needs to be further improved.
The next generation of models will therefore aim at determining radiative
accelerations with higher precision (see Sect.\,\ref{subsec:perspective}).
\par
One independent result, however, might be worth considering in this
context: Detailed parallax and gravitational redshift measurements with HST
of the binary system Feige\,24 \citep{benedict:00,vennes:00} also yielded a
higher surface gravity for the WD component than several optical analyses
($\log\,g$=7.2-7.5). The consistency of our result ($\log\,g$=7.7) with
\citeauthor{vennes:00}'s ($\log\,g$=7.5-7.7) might be interpreted as a hint that our
current results do not suffer from an inaccurate $g_{rad}$ determination in
the models after all. Instead, analyses with homogeneous models would have
been biased due to systematic errors introduced through the neglect of
stratification. But even though this may sound plausible, it is too early
to call such a statement anything more than a suspicion.

\subsection{Conclusions and perspective}
\label{subsec:perspective}

In this paper, we have presented our chemically stratified model atmospheres
describing the chemical composition through the equilibrium between
gravitational settling and radiative acceleration self-consistently. The
application to 26 EUVE spectra of hot DA white dwarfs has revealed
an overall good agreement, demonstrating the potential of these new models. 
\par
In upcoming papers, we will check the derived parameters through a
re-analysis of the Balmer lines and of the metal lines observable in UV
spectra. The analysis of optical spectra with our new stratified models
should allow to check for systematic error as compared to previous analyses
with chemically homogeneous models. This will also be an important test for
the consistency of atmospheric parameters derived from different parts of
the spectrum, which has often posed a problem in the past, and to which
we hope the new type of models will ultimately be a solution. Only this
test will show on how a firm ground the parameters presented here stand.
Likewise, the determination of the interstellar column densities has to be
regarded as preliminary, since these cannot be determined independently
from the theoretical spectra.
\par
The current models are only a first step, in the future we will be able to
present an improved set of stratified model atmospheres. A ``next
generation'' version of our stellar atmosphere code itself as well as a
more efficient coupling between the computation of the model structure and
the chemical stratification will allow to include much more detailed
model atoms and at the same time many more chemical elements. This
influences and is crucial for the accuracy of the calculation of the
radiative acceleration. By evaluating the scale of the deviation in
comparison to the current results it should then become clear how
trustworthy the latter are.

\begin{acknowledgements}
  The authors would like to thank Klaus Werner (T\"ubingen) for useful
  comments and discussions. We also thank Detlev Koester (Kiel) who
  provided part of the source code for the handling of the ISM absorption.
  The referee has provided very valuable comments to help improving this
  paper.  All model atmospheres have been calculated on CRAY machines of
  the Rechenzentrum der Universit\"at Kiel. This work is supported by the
  Deutsche Forschungsgemeinschaft under grant DR 281/13-1.
\end{acknowledgements}

\end{document}